# Modeling and Uncertainty Analysis of Groundwater Level Using Six Evolutionary Optimization Algorithms Hybridized with ANFIS, SVM, and ANN

**Akram Seifi [1],\*, Mohammad Ehteram [2], Vijay P. Singh [3] and Amir Mosavi [4],\***

[1] Department of Water Science & Engineering, Vali-e-Asr University of Rafsanjan, Rafsanjan, Iran
[2] Department of Water Engineering and Hydraulic Structures, Faculty of Civil Engineering, Semnan University, Semnan 35131-19111, Iran.
[3] Department of Biological and Agricultural Engineering & Zachry Department of Civil Engineering Texas A&M University College Station, Texas, TX 77843-2117, USA; vsingh@tamu.edu
[4] Institute of Structural Mechanics, Bauhaus-Universität Weimar, 99423 Weimar, Germany.
\* Correspondence: a.seifi@vru.ac.ir, amir.mosavi@uni-weimar.de

**Abstract:** In the present study, six meta-heuristic schemes are hybridized with artificial neural network (ANN), adaptive neuro-fuzzy interface system (ANFIS), and support vector machine (SVM), to predict monthly groundwater level (GWL), evaluate uncertainty analysis of predictions and spatial variation analysis. The six schemes, including grasshopper optimization algorithm (GOA), cat swarm optimization (CSO), weed algorithm (WA), genetic algorithm (GA), krill algorithm (KA), and particle swarm optimization (PSO), were used to hybridize for improving the performance of ANN, SVM, and ANFIS models. Groundwater level (GWL) data of Ardebil plain (Iran) for a period of 144 months were selected to evaluate the hybrid models. The pre-processing technique of principal component analysis (PCA) was applied to reduce input combinations from monthly time series up to 12-month prediction intervals. The results showed that the ANFIS-GOA was superior to the other hybrid models for predicting GWL in the first piezometer (RMSE:1.21, MAE:0.878, NSE:0.93, PBIAS:0.15, R2:0.93), second piezometer (RMSE:1.22, MAE:0.881, NSE:0.92, PBIAS:0.17, $R^2$:0.94), and third piezometer (RMSE:1.23, MAE:0.911, NSE:0.91, PBIAS:0.19, $R^2$:0.94) in the testing stage. The performance of hybrid models with optimization algorithms was far better than that of classical ANN, ANFIS, and SVM models without hybridization. The percent of improvements in the ANFIS-GOA versus standalone ANFIS in piezometer 10 were 14.4%, 3%, 17.8%, and 181% for RMSE, MAE, NSE, and PBIAS in training stage and 40.7%, 55%, 25%, and 132% in testing stage, respectively. The improvements for piezometer 6 in train step were 15%, 4%, 13%, and 208% and in test step were 33%, 44.6%, 16.3%, and 173%, respectively, that clearly confirm the superiority of developed hybridization schemes in GWL modelling. Uncertainty analysis showed that ANFIS-GOA and SVM had, respectively, the best and worst performances among other models. In general, GOA enhanced the accuracy of the ANFIS, ANN, and SVM models.

## 1. Introduction

One of the most important sources of water supply for industrial, drinking, and irrigation purposes is groundwater (GW). GW has a significant role in economic development, environmental management, and ecosystem sustainability [1,2]. However, in recent years undue exploitation has caused a tremendous pressure on GW resources, resulting in GW crisis [3]. As a result, the GW level (GWL) in different regions of the world has been decreasing rapidly. Further, widespread pollution of surface water is severely affecting GW. A decrease in GWL can also be caused by climate factors and can lead to a number of eco-environmental problems [4]. For proper water resources management, particularly effective utilization and sustainable management of groundwater resources, accurate and reliable prediction of GWL is essential [5,6]. Thus, it is necessary to predict the Ardebil groundwater level for water resources management. Mathematical models incorporating GW dynamics are applied to predict GWL for optimizing groundwater use, optimal management, and development of conservation plans [5,7]. Since such models are costly, time-consuming, and data-intensive, their use in practice is limited because of data-scarcity [8,9]. In such cases, when geological and hydro-geological data are insufficient, soft computing models become an attractive option [10]. Artificial neural network (ANN), adaptive neuro-fuzzy interface (ANFIS), genetic programming (GP), support vector machine (SVM), and decision tree models are among the important soft computing models that are suited for modeling dynamic and uncertain nonlinear systems [7].

Recently, soft computing models have been widely used worldwide to predict GWL. Jalal Kameli et al. [11] evaluated neuro-fuzzy (NF) and ANN models to estimate GWL using rainfall, air temperature, and GWLs in neighboring wells, and showed that the NF model performed better than the ANN model. Identifying the lag time of time series for observed rainfall by correlation analysis, Trichakis et al. [12] used the ANN model to predict GWL and found the ANN model to be useful to model Karst aquifers that are difficult to simulate using numerical models. Using evaporation, rainfall, and water levels in observation levels as input, Fallah-Mehdipour et al. [13] applied the ANFIS and genetic programming models for predicting GWL and showed that GP decreased the value of mean root square error (RMSE) compared to the RMSE by the ANFIS. Moosavi et al. [14] evaluated the ANN, ANFIS-wavelet, and ANN-wavelet models and showed that predicted GWL was more accurate for 1 and 2 months ahead than for 3 and 4 months ahead. Predicting GWL in the Bastam plain by ANFIS and ANN models in Emamgholizadeh et al. [15] study confirmed that if the water shortage of the aquifer remained equal to the pumping rate of water from wells, the minimum reduction of GWL occurred. Suryanarayana et al. [16] proposed a hybrid model integrating the SVM model with the wavelet transform and indicated that the SVM-wavelet model was more accurate in predicting GWL. Using rainfall, pan evaporation, and river stage as input, Mohanty et al. [17] indicated that the ANN model was better using shorter lead times for GWL predictions than the larger lead times. Yoon et al. [18] demonstrated that the SVM model was superior to the ANN model in predicting GWL. Zho et al. [19] found that the wavelet-SVM model was better than the wavelet-ANN model for modelling GWL. Comparing ANN and autoregressive integrated moving average (ARIMA), Choubin and Malekian [20] showed that the ARIMA model was more accurate than ANN in modelling GWL. Das et al. [21] found ANFIS to be better than ANN for predicting GWL.

Literature review shows that although soft computing models are capable for predicting groundwater level, they have weaknesses and uncertainties [22]. The ANN models have different parameters, such as weight connections, bias, and need training algorithms to fine-tune their parameters. ANFIS and SVM models have nonlinear and linear parameters and use different kinds of training algorithms, such as backpropagation algorithm, descent gradient method, etc. However, the standard training algorithms have two major defects: slow convergence and getting trapped in local optima [22]. Recently, nature-based optimization algorithms have been developed for finding the appropriate values of model parameters to improve ANN, ANFIS, and SVM models. Jalalkamali and Jalalkamali [23] applied a hybrid model of ANN and genetic algorithm (ANN-GA) to find the best number of neutrons for the hidden layer and predict GWL in an individual well. Mathur [24] applied hybrid SVM-PSO (particle swarm optimization) model for predicting GWL in Rentachintala region of Andhra Pradesh, India, where optimal parameters of SVM were determined using PSO. Results showed that SVM-PSO was more accurate than the ANN, ANFIS, and ARMA models. Hosseini et al. [25] hybridized ANN and ant colony optimization (ACO) to predict the GWL in

Shabestar plain, Iran, and found that the hybrid ANN-ACO model reduced overtraining errors. Zare and Koch [26] demonstrated that the hybridized wavelet-ANFIS model was superior in modelling GWL to other regression models. Balavalikar et al. [27] found that the hybrid ANN-PSO model was better in predicting monthly GWL of Udupi district, India, than the classical ANN model. Malekzadeh et al. [28] evaluated ANN, wavelet extreme machine learning (WEML), SVM, wavelet-SVM, and wavelet-ANN for predicting GWL, and concluded that WEML was more accurate. These studies reveal that hybrid models are more accurate and efficient than single models in predicting GWL and it is inferred from these studies that meta-heuristic optimization algorithms are superior to the classical ones, but require uncertainty analysis for artificial intelligence models.

New hybrid intelligent optimization models can be regarded as appropriate alternative methods with an acceptable range of error for predicting GWL. Among the nature-inspired optimization algorithms, the grasshopper optimization algorithm (GOA) is a novel and robust meta-heuristic method that mimics the swarming behavior of grasshoppers in nature. The GOA is a multi-solution-based algorithm during the optimization process to avoid higher local optima and has high convergence ability toward the optimum [29]. It has different functions than other optimization algorithms that enable it to find the best optimal solution in the search space with high probability. Therefore, this algorithm escapes from local optima and finds the global optimum in the search space. This capability is considered as an advantage of GOA [30] and as reason for the selection of GOA for the current study. Several researchers used GOA for monthly river flow [31], soil compression coefficient [32], coefficients of sediment rating curve [33], and concrete slump [34], but the uncertainty analysis and GWL modeling has not yet been studied.

These models have some drawbacks in the previous studies that are addressed in the current paper. These models are robust tools for modeling many of the nonlinear hydrologic processes such as rainfall-runoff, stream flow, and ground-water level. Despite the wide application of soft computing models, few studies have investigated the capability of novel optimization algorithms, such as GOA integrated with typical predictive methods, for GWL prediction, uncertainty evaluation, and spatial variation modeling. The main problem in developing these models is the using of an appropriate training procedure. Especially, AI tend to be very data intensive in training stage, and there appears to be no established methodology for design and successful implementation of training procedure and error minimizations. Therefore, there are still some questions about AI tools that must be further studied, and important aspects such as local trapping, uncertainty analysis of results, uncertainty due to meta-heuristic optimization algorithms in training, spatial changes modelling with hybrid models must be explored further. Based on the best knowledge of the authors, no published papers exist that evaluate the uncertainty of different meta-heuristic optimizations for groundwater level prediction in hybridization with ANN, ANFIS, and SVM. The main contribution and novelty of the present study is comparative uncertainty analysis of the novel hybrid models, spatial changes modelling by considering PCA as appropriate input selection in regard to uncertainty results. Despite the wide application of soft computing models, few studies have investigated the capability of novel optimization algorithms, such as GOA integrated with typical predictive methods, for GWL prediction, uncertainty evaluation, and spatial variation modeling. The state-of-art models, including ANN, ANFIS, and SVM, have been employed to predict GWL, but these models are easily trapped in local optima and often need longer training times. Hence, the main contribution of this study is to develop and to assess the applicability of hybrid ANFIS-GOA, SVM-GOA, and ANN-GOA models for predicting monthly GWL and uncertainty of results in Ardabil basin in Iran. Application of GOA method integrated with ANN, ANFIS, and SVM models is useful to search the best numerical weights of neurons and bias values. The other objectives of this paper were to (1) compare the GOA with different optimization algorithms of particle swarm (PSO), weed algorithm (WA), cat algorithm (CA), and genetic algorithm (GA); (2) evaluate the uncertainty of the hybridized models for predicting monthly GWL; (3) use principal component analysis to select the appropriate input combinations from time-series data up to 12-month lag; (4) modeling spatial variation of GWL by using hybrid intelligence models results in geospatial analysis.

## 2. Materials and Methods

*2.1. Case Study and Data*

The Ardebil plain, with the area of 990 km², is located in the northwest of Iran between latitudes 38′3° and 38′27 and the longitudes of 47′55° and 48′20° (Figure 1). The average annual rainfall is 304 mm. The hottest month in this plain is May and the driest month is July. The average annual temperature is 9 °C. In Ardebil plain, groundwater supplies water for drinking, agricultural, and industrial purposes. There is a negative balance of about 550 million m³ in the Ardebil aquifer. The GWL decreases by 20–30 cm per year, which is the fastest decline. The Ardebil plain has 89 villages, that use groundwater for agricultural uses. The current condition of the GWL in the Ardebil plain has negative impacts on the farmers as its main users. In this study, the following parameters were used as the input to the hybrid ANN, ANFIS, and SVM models. Then, the principal component analysis was used to select the best input combination up to 12-month lag.

$$H(t) = f[H(t-1), H(t-2), H(t-3), \ldots H(t-12)] \quad (1)$$

where, $H(t)$ is the GWL at month t, $H(t-1)$ is the 1-month lagged H, $H(t-2)$ is the 2-month lagged H, $H(t-3)$ is the 3-month lagged H, and $H(t-12)$ is the 12-month lagged H. The data of 140 months (2000 (January)–2012 (September)) were selected for the current study. A total of 20% of the data set was used for testing, and 80% of the data set was used for the training, that were selected randomly. Nine observed wells (wells 6, 9, 10, 24, 11, 4, 7, 8, and 1) were used to provide the spatiotemporal variation of GWL for different months. Each piezometer had 140 monthly data points. The measurements were made one time during each month.

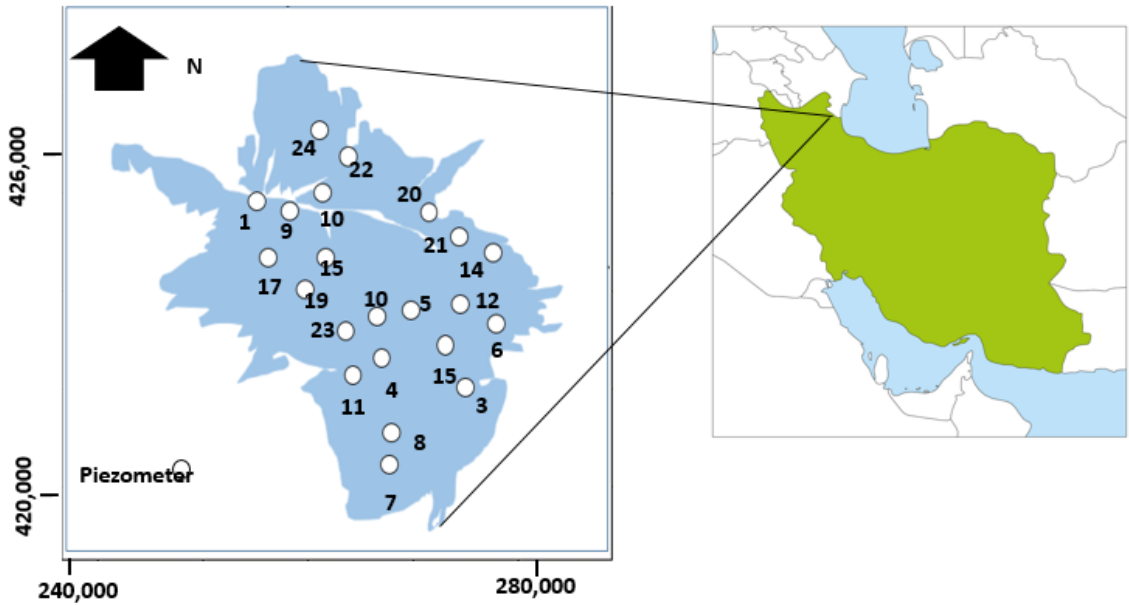

**Figure 1.** Location of Ardebil Plain as the case study.

*2.2. ANFIS Model*

The ANFIS model uses fuzzy interface systems which use fuzzy if-then rules to construct a predictive model. The ANFIS model has been widely used for predicting rainfall [33], temperature [34], runoff [35], evaporation [36], and sediment load [37]. Figure 1 shows the structure of the ANFIS model in the framework of the study. The square nodes and circle nodes show the adaptive and fixed nodes, respectively. The ANFIS model has five layers [38]. (1) The inputs are fuzzified in the first layer whose nodes are constant. The membership grade of inputs is the output of the first layer:

$$\begin{aligned} o_i^1 &= u_{Ai}(x), i = 1,2.. \\ o_i^1 &= u_{B_{i-2}}(y), i = 3,4,.. \end{aligned} \quad (2)$$

where, $o_i^1$ is the output of the first layer, $u_{A_i}(x)$ and $u_{B_{i-2}}(y)$ are the fuzzy membership functions for the fuzzy set $A_i$ and $B_{i-2}$, respectively. The bell-shaped member function is selected for the current study due to its smoothness and concise notation:

$$u_{A_i}(x) = \frac{1}{1 + \left[\left(\frac{x - c_i}{a_i}\right)^2\right]^{b_i}}, i = 1,2,.. \tag{3}$$

where $a$, $b$, and $c$ are the premise parameters (training algorithms obtain these parameters).

(2) The nodes of the second layer are labelled with M, which shows that they carry out a simple multiplier function. The fuzzy strengths $\omega_i$ of each rule are the output of the second layer:

$$o_i^2 = \omega_i = u_{A_i}(x)u_{B_i}(y), i = 1,2.., \tag{4}$$

(3) The nodes of the third layer are also fixed. The fuzzy strengths from the previous layer are normalized in the third layer. The sum of weight functions is used to compute the normalization factor. The normalized fuzzy strengths are the output of the third layer:

$$o_i^3 = \bar{\omega}_i = \frac{\omega_i}{\sum_{i=1}^{2} \omega_i} \tag{5}$$

(4) The nodes of the fourth layer are adaptive and its outputs are computed as:

$$o_i^4 = \bar{\omega}_i z_i = \bar{\omega}_i(p_i + q_i y + r_i), i = 1,2.., \tag{6}$$

where, $p_i$, $q_i$, and $r_i$ are the consequent parameters.

(5) The output in the fifth layer is labelled with S. A fixed node is observed in this layer. This layer computes the total summation of all the incoming signals:

$$o_i^5 = z = \sum_{i=1}^{2} \bar{\omega}_i z_i = \frac{\sum_{i=1}^{2} \omega_i z_i}{\sum_{i=1}^{2} \omega_i} \tag{7}$$

In the classical training approach, a combination of the least square and gradient descent methods is commonly used as a hybrid learning algorithm to adjust the parameters of the ANFIS model. The consequent parameters of ANFIS model are updated by applying the least square method in the forward pass. Additionally, in the backward pass, the gradient descent method is used for updating the premise parameters. In the hybridized schemes, tuning and adjusting the consequent and premise parameters are determined by the optimization algorithms as the hybrid training scheme.

*2.3. ANN Model*

The artificial neural network uses behavioral patterns to provide a framework for modeling mechanisms. It consists of three layers: input, hidden, and output layers, and includes the processing units named neurons which are arranged in several layers [39]. The connection weights link the neurons of preceding layers to the neurons of the following layers. The output of the middle layer (hidden layer) is used as the input to the following layer. The input data is received by the input layer, while the last layer generates the final output of the ANN model. The middle layers receive and transmit the input data to the connected nodes in the following layers. The weighted sum of inputs is used by the hidden neurons to produce the intermediate output. The ANN model uses the activation functions to compute the outputs of the hidden and output neurons. It uses the bias values to set the output along with the weighted sum of inputs to the neuron. The process of ANN modelling has two major levels: (1) preparing the network structure, and (2) adjustment of the weights of connections. The literature review indicates that the backpropagation training algorithm is wildly used in different fields, such as water engineering [40]. First, the output of the ANN model is obtained as a response of the ANN model. In the next level, the error between observed and estimated values is minimized to find the weights of the model. If the output is different from the observed value, the modification of weights and biases will start to decrease the error values. However, the backpropagation algorithm has a slow convergence rate and to overcome its inherent weakness the meta-heuristic optimization algorithms are used in the present study. Figure 1 shows the structure of the ANN model and its hybridization with intelligence algorithms.

*2.4. SVM Model*

The SVM model has been widely used for predicting solar radiation [41], rainfall [42], landslides [43], and drought [44]. In the SVM model, the input data are divided into testing and training samples. The selected input vector (training sample) is mapped into a high-dimensional feature space. Then, the optimal decision function is generated [44]. Equation (7) shows the regression estimation function of the SVM model:

$$f(x) = W^T \phi(x) + b \tag{8}$$

where, $\phi(x)$ is the nonlinear mapping function for mapping sample data (x) into an m-dimensional feature vector, $b$ is the bias, and $W^T$ is the weight vector of the independent function. $W^T$ and $b$ are computed by minimizing the following function:

$$D(f) = \frac{1}{2}\|w\|^2 + \frac{C}{n}\sum_{j=1}^{n} R_\varepsilon[y_j, f(x_j)] \tag{9}$$

where, $D(f)$ is the generalized optimal function, $\|w\|^2$ is the complexity of the model, $C$ is the penalty parameter, and $R_\varepsilon$ is the error control function of $\varepsilon$. Thus, the optimization problem is defined as follows:

$$\begin{aligned} \min Q\,(W,\xi) &= \frac{1}{2}\|W\|^2 + C\sum_{j=1}^{n}\xi_j + \xi_j^n \\ W^T\phi(x_j) + b - y_j &\leq \varepsilon + \xi_j \\ y_j - W^T\phi(x_j) - b &\leq \varepsilon + \xi^*_j \\ \xi_j &\geq 0, \xi^*_j \geq 0, j = 1,2,\dots,n \end{aligned} \tag{10}$$

where, $\xi_j$ and $\xi^*_j$ are the relation factors. Adjusting the partial derivatives of $W$, $b$, $\xi_j$, and $\xi^*_j$ to 0 and using the Lagrangian equation, an optimization problem can be formulated as follows:

$$\begin{aligned} L(W,a,b,\varepsilon,y) &= \min \frac{1}{2}\sum_{j=1}^{n}(a_r - a_r^*)^T H_{r,j*} \\ a_r - a_r^* + \varepsilon\sum_{j=1}^{n}(a_r - a_r^*) &+ \sum_{j=1}^{n} y_r(a_r - a_r^*) \\ \sum_{r=1}^{n}(a_r - a_r^*) &= 0, (0 \leq a_r, a_r^* \leq C) \\ H_{r,j} &= K(x,x_j) = \phi(x_r)^T \phi(x_j), (r = 1,2,\dots,n) \end{aligned} \tag{11}$$

where, $K(x, x_j)$ is the kernel function. The most popular kernel function is the radial basis function:

$$K(x, x_j) = exp\left(-\frac{|x - x_j|^2}{2\gamma^2}\right) \tag{12}$$

where, $\gamma$ is the radial basis function parameter. The SVM based model uses the grid search algorithm (GS) to find the optimal value of parameters C and $\gamma$. Specifically, a set of initial values is chosen for both parameters $\gamma$ and C. To select $\gamma$ and C using cross-validation, the available data are divided into k subsets. One subset is regarded as testing data and then assessed using the remaining k-1 training subsets. Then, the cross-validation error is computed using the split error for the SVM model using different values of C and $\gamma$. Various combination of parameters C and $\gamma$ are evaluated and the one yielding the lowest cross-validation error is chosen and used to train the SVM model for the whole dataset. The structure of the SVM model is shown in Figure 2.

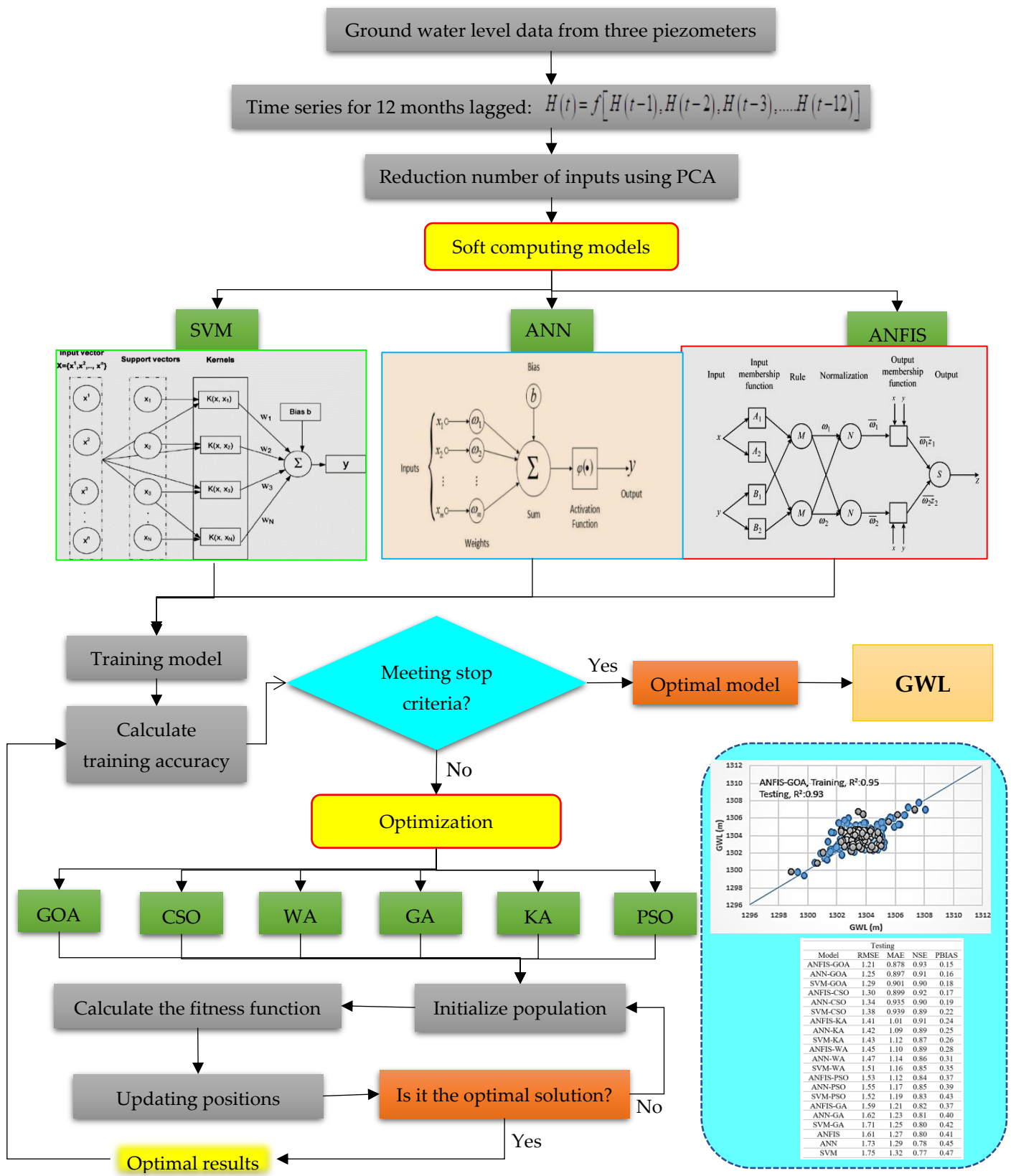

**Figure 2.** Developed methodology framework for modeling groundwater level time series.

*2.5. Optimization Algorithms*

2.5.1. Grasshoppers Optimization Algorithm (GOA)

Grasshoppers are regarded as pests because they damage agricultural crops. They are a group of insects that can generate large insect swarms. The mathematical function to investigate the swarming behavior of grasshoppers is demonstrated with the following equation [45]:

$$X_i = S_i + G_i + A_i \tag{13}$$

where, $X_i$ is the position of the $i^{th}$ grasshopper, $S_i$ is the classical interaction, $G_i$ is the gravity force on the $i^{th}$ grasshopper, and $A_i$ is the wind advection. The classical interaction is simulated as follows:

$$S_i = \sum_{j=1}^{N} s(d_{ij}) \hat{d}_{ij} \tag{14}$$

where, $d_{ij}$ is the distance between the $i^{th}$ and $j^{th}$ grasshoppers, and $s$ is a function for the definition of the strength of social forces.

$$\begin{aligned} d_{ij} &= |x_i - x_j| \\ \hat{d}_{ij} &= \frac{x_j - x_i}{d_{ij}} \end{aligned} \tag{15}$$

The function $s$ is computed as follows:

$$s(r) = f e^{-\frac{r}{l}} - e^{-r} \tag{16}$$

where, $f$ is the intensity of attraction, and $l$ is the attractive length scale. The distance between grasshoppers ranges between 0 and 15. Repulsion is observed in the interval [0 2.079]. The grasshoppers enter the comfort zone if they are far from 2.079 units from other grasshoppers. $G$ component is computed as follows:

$$G_i = -g\hat{e}_g \tag{17}$$

where, $g$ is the gravitational constant and $\hat{e}_g$ is a unity vector towards the center of the earth. The $A$ parameter is computed as follows:

$$A_i = u\hat{e}_w \tag{18}$$

where, $u$ is a constant drift and $\hat{e}_w$ is a unit vector in the direction of the wind. Finally, the new position of a grasshopper is computed using its common position, the food source position, and the position of all other grasshoppers:

$$X_i = \sum_{\substack{j=1 \\ j \neq i}}^{N} s(|x_j - x_i|) \frac{x_j - x_i}{d_{ij}} - g\hat{e}_g + u\hat{e}_w \tag{19}$$

where, $N$ is the number of grasshoppers. However, Equation (18) cannot be directly used for optimization because grasshoppers do not converge to a specified point. Thus, a corrected equation is used to update the grasshopper's position:

$$X_i^d = c\left[\sum_{\substack{j=1 \\ j \neq i}}^{N} c\frac{ub_d - lb_d}{2} s(|x_j^d - x_i^d|) \frac{x_j - x_i}{d_{ij}}\right] + \hat{T}_d \tag{20}$$

where, $ub$ is the upper bound; $lb_d$ is the lower bound; $\hat{T}_d$ is the value of the $D_{th}$ dimension in the target space (optimal solution found so far); and $c$ is a decreasing coefficient to shrink the comfort zone, repulsion zone, and attraction zone. Figure 2 shows the flowchart of GOA.

2.5.2. Weed Algorithm (WA)

Weeds have a very adaptive nature that converts them to undesirable plants in agriculture. Figure 3 shows the flowchart of the WA algorithm [46]. The WA starts with initializing a random population of weeds in the search space. A predefined number of weeds are randomly distributed over the entire dimensional space, indicated as a solution space. The fitness of weeds is assessed by considering its fitness function to optimize the problem. Each agent of the current population can produce some seeds via a predefined region considering its own location. In this way, the number of produced seeds relies on its fitness function in the population regarding the best and worst solutions, as observed in Figure 3. The number of seeds is computed as follows [46]:

$$Number (of) seed (around) weed_i = \frac{F_i - F_{worst}}{F_{best} - F_{worst}}(Smin_{max} + S_{min}) \quad (21)$$

where, $F_{worst}$ is the worst fitness function, $F_{best}$ is the best fitness function, $S_{min}$ is the minimum number of seeds, $S_{max}$ is the maximum number of seeds, and $F_i$ is $i^{th}$ fitness function. The distribution of seeds is random over the search space and is based on the standard deviation $\sigma_i$ and zero mean. The standard deviation of the distribution of seeds varies as follows:

$$\sigma_{cur} = \frac{(iter_{max}()^n)}{(iter_{max}()^n)(\sigma_{init} - \sigma_{final}) + \sigma_{final}} \quad (22)$$

where, $iter_{max}$ is the maximum number of iterations, $\sigma_{cur}$ is the standard deviation at the current iteration, $\sigma_{final}$ is the final value of standard deviation, $\sigma_{init}$ is the predefined initial value of standard deviation, and $n$ is the nonlinear modulation index. Seeds are produced by each weed and then are distributed over the space. The competitive exclusion is the final level in the WA. If a weed does not generate seeds, it will be extinct. If all the weeds generate seeds, the number of weeds increases exponentially. Therefore, the number of seeds is limited to the maximum value ($P_{max}$). The weeds with better fitness function are allowed to reproduce. Weeds with worse fitness function are removed (see figure 4).

```
1: Objective function f(x), x = (x1, x2, ….., xdim), dim= no. of dimensions
2: Generate initial population of n grasshoppers xi= (i=1, 2, ….., n)
3: Calculate fitness of each grasshopper
4: T = the best search agent
5: while stopping criteria not met do
6: Update c1 using equation (20)
7: for each grasshopper gh in population do
8: Normalize the distances between grasshoppers in [1,4]
9: Update the position of the grosshopers by Eq. (19)
10: If required, update bounds of gh
11: end for
12: If there is a better solution, update T
13: end while
14: Output the T.
```

**Figure 3.** The flowchart of grasshopper optimization algorithm (GOA) [30].

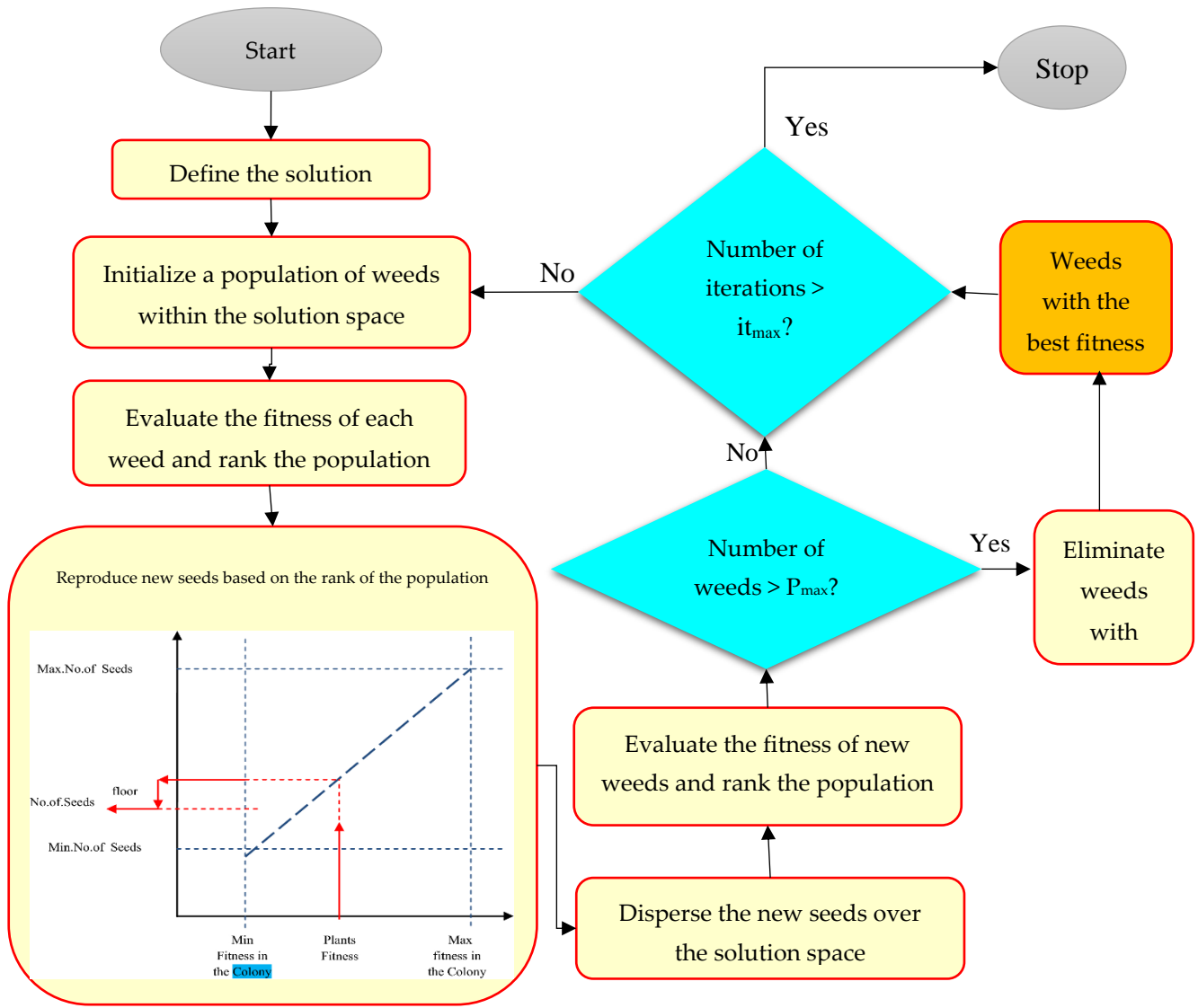

**Figure 4.** The flowchart of weed algorithm (WA) [46].

2.5.3. Cat Swarm Optimization (CSO)

Recently, CSO has gained popularity among other optimization algorithms because of its exploration ability and is widely used in different fields, such as wireless sensor networks [47], robotics [48], data clustering [49], and dynamic multi-objective algorithms [50]. Chu et al. (2006) introduced the cat swarm algorithm [51]. Figure 5 shows the flowchart of CSO. The CSO uses hunting and resting skills for optimization. First, the initial population of cats is initialized randomly. The seeking mode and tracing mode are two important operation modes in the CSO model. The seeking mode demonstrates the resting ability of cats which change their position and remain alert. This mode is regarded as a local search for the solutions. The seeking memory pool (SMP), the seeking range of selected dimension (SRD), and counts of dimension to change (CDS) affect the cat's behavior. The number of duplicate cats is denoted by SMP. CDC shows that the dimensions are to be mutated and SRD denotes change value of chosen dimensions. In the seeking mode, most of the cat's time is in the resting time, even though they remain alert [52]. The seeking mode includes the following levels:

- Generate replicas of the cats as per SMP.
- The position of each copy is updated as follows:

$$x_{k,d} = \begin{bmatrix}(1 + (2 \times rand - 1) * SRD) * x_{j,d} \leftarrow if\,(D) \in N \\ x_{j,d} \leftarrow otherwise\end{bmatrix} \quad (23)$$

where, $x_{k,d}$ is the position of the $k^{th}$ cat in the $d^{th}$ dimension (new position of the cat), $rand$ is the random number, $N$ is the number of cats, $D$ is the number of dimensions, and $x_{j,d}$ is the position of $j^{th}$ cat in the $d$ dimension (old position of the cat).

- Compute the objective function for all copies and choose the best objective function value ($x_{best}$) of the cat.
- Substitute $x_{j,p}$ with the best cat if the $x_{best}$ is better than $x_{j,p}$ in terms of the objective function value.

The hunting skill of cats is represented by the tracing mode. Cats trace the objectives with high energy by changing their locations with their own velocities. The velocity is updated as follows:

$$v_{j,d,new} = \omega \times v_{j,d} + r_1 \times c_1 \times (x_{best,d} - x_{j,d}) \quad (24)$$

where, $\omega$ is the inertia weight, $c_1$ is a constant, and $v_{j,d}$ is the velocity of $j^{th}$ cat in the $d$ dimension, and $v_{j,d,new}$ is the new velocity of the $j^{th}$ cat. The position of cats in the tracing mode is updated as follows:

$$x_{j,d} = x_{j,d} + v_{jd} \quad (25)$$

where, $x_{j,d_{new}}$ is the $j^{th}$ position of the $k^{th}$ cat in the $d^{th}$ dimension (new position of the cat).

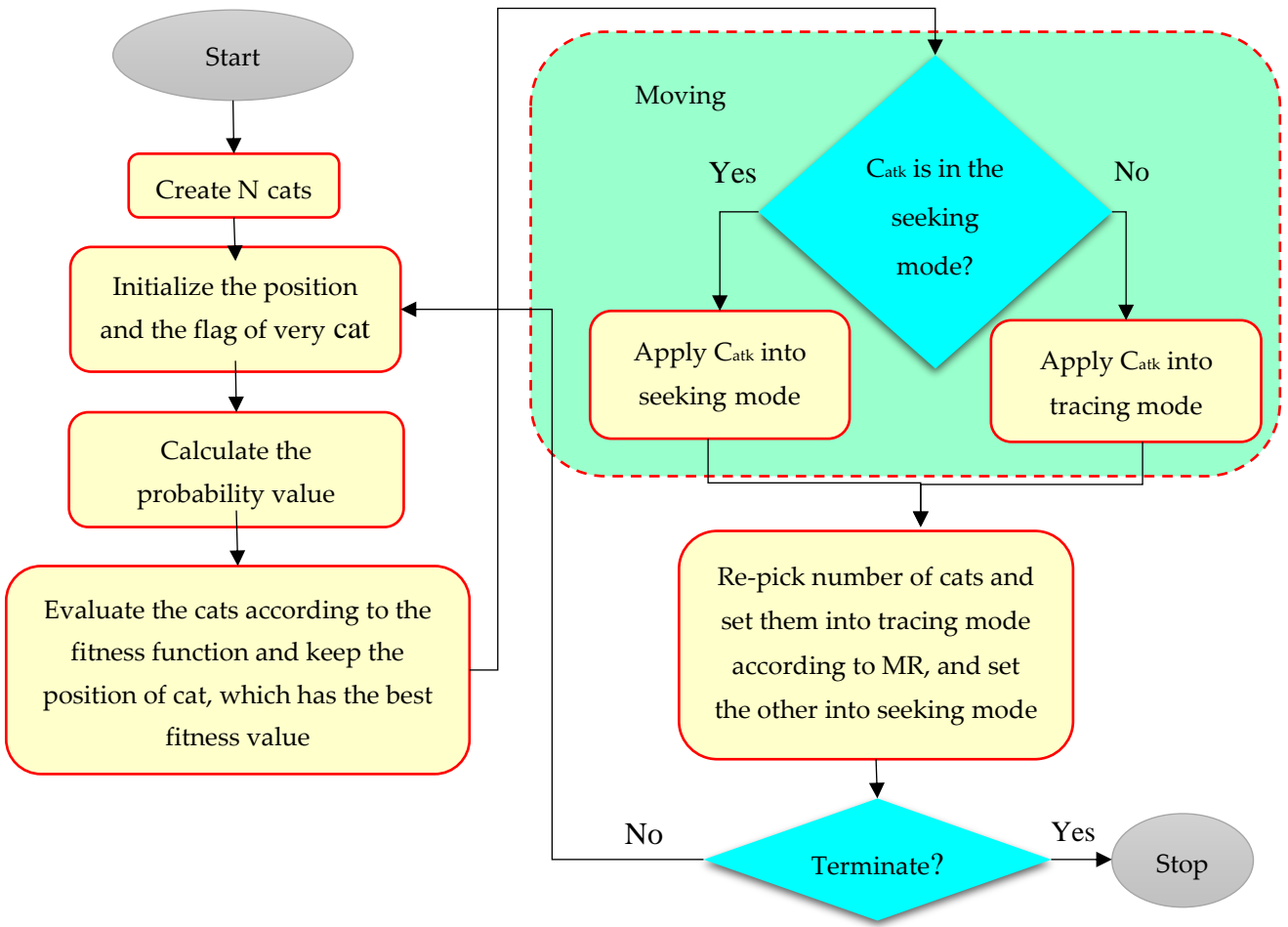

**Figure 5.** Flowchart of cat swarm optimization (CSO) for the optimization problems [49].

2.5.4. Particle Swarm Optimization (PSO)

In PSO, a set of particles that are generated randomly search the best adjacent solutions for optimization. The updating equations for the new position and velocity of particles are written as [53]:

$$x_{id}(t+1) = x_{id}(t) + v_{id}(t+1) \qquad (26)$$

$$v_{id}(t+1) = \psi * v_{id}(t) + r_1 * c_1 * [p_{id}(t) - x_{id}(t)] + r_2 * c_2 * [g_d(t) - x_{id}(t)] \qquad (27)$$

where, $d$ is the number of dominions; $\psi$ is the inertia weight; $r_1$ and $r_2$ are the random values; $c_1$ and $c_2$ are the acceleration coefficients; $g_d(t)$ is the global best position obtained by neighbors; and $p_{id}$ is the personal best position.

The particles find the solutions of optimization problems by adjusting the position and velocity of particles. The main advantages of PSO are easy implementation and computational efficiency.

### 2.5.5. Genetic Algorithm (GA)

Genetic algorithm is one of the most popular algorithms that is extensively applied for optimization problems. Each chromosome in GA is a candidate solution [19]. The genes of chromosomes simulate the variables of optimization. First, the initial population of chromosomes is randomly initialized for optimization and the selection operator is used to select the best chromosomes for the production of the next generation. The chromosomes with better fitness values have a great chance of being chosen by the selection operator. The crossover operator is used to exchange genes between two chromosomes for producing new solutions. Finally, the mutation operator is used to cause changes in the genes. The mutation operator is applied to the chromosomes of new genes to generate different solutions with new genes. If the convergence criteria are satisfied, the algorithm stops; otherwise, the algorithm runs again. The drawback of GA shows that GA requires a high number of iterations [20].

### 2.5.6. Krill Herd Algorithm (KHA)

Gandomi and Alavi [54] introduced the KHA using the krill's behavior in nature [54]. The KHA is widely used in different fields, such as text document clustering analysis [55] and structural seismic reliability [56]. The KHA acts, based on three main concepts: (1) mutation-induced, (2) foraging mutation, and (3) physical diffusion. The following formulation uses the three behaviors mentioned above [54]:

$$\frac{dY_i}{dt} = N_i + F_i + D_i \qquad (28)$$

where, $Y$ is the location of the $i^{th}$ krill, $N_i$ is the motion induced by another krill, $F_i$ is the foraging motion, and $D_i$ is the physical diffusion of the $i^{th}$ krill. Equation (28) describes the motion-induced by another individual krill.

$$N_{new,i} = N(\alpha_{local,i} + \alpha_{target,i})_{n_{old,i}}{}_{max} \qquad (29)$$

where, $N_{max}$ is the maximum induced speed, $\alpha_{local,i}$ is the neighbor's local effect, $\alpha_{target,i}$ is the krill's target direction, $\omega_n$ is the inertia weight of induced motion, and $N_{old,i}$ is the old motion-induced for the $i^{th}$ individual krill. The foraging motion can be formulated as:

$$F_i = V_f(\beta_{food,i} + \beta_{best,i}) + \omega_f F_{old,i} \qquad (30)$$

where, $V_f$ is the foraging speed, $\beta_{food,i}$ is the food attractive, $\beta_{best,i}$ is the effect of the best fitness of the $i^{th}$ krill, and $F_{old,i}$ is the last foraging motion. The diffusion can be computed as:

$$D_i = D_{max} \qquad (31)$$

where, $D_{max}$ is the maximum diffusion speed, and $\delta$ is the random direction.

Finally, the position of a krill is computed as follows:

$$X_{new,i} = X_{old,i} + \Delta t \frac{dX_i}{dt} \tag{32}$$

where, $X_{new,i}$ is the value of the next individual krill location, and $X_{old,i}$ represents the current position of solution number I, and $\Delta t$ is the essential constant. Figure 5 shows the flowchart of the krill algorithm.

*2.6. Principal Component Analysis (PCA)*

PCA is a statistical orthogonal transformation to obtain a set of values of linearly uncorrelated (principal components) from a set of observations. When the user has the number of inputs but he cannot identify the appropriate inputs, the PCA is used to reduce the number of inputs. The final data set should be able to demonstrate most of the variance of the original input data by creating a variable reduction [57]. PCA can be explained, based on the following equation [57]:

$$Z_i = a_{i1} + a_{i2} + \ldots + a_{ip} x_p \tag{33}$$

where, $Z_i$ shows the principal component, $a_{ip}$ is the related eigenvector, and $x_i$ is the input variable. The information is obtained by solving Equation (34):

$$|R - \lambda I| = 0 \tag{34}$$

where, $R$ is the variance-covariance matrix, $I$ is the unit matrix, and $\lambda$ is the eigenvalues.

*2.7. Taguchi Model*

The random parameters of optimization algorithms are the most important parameters affecting the outputs of the optimization algorithms. Thus, determining the appropriate values of random parameters is necessary to construct the optimization models. The Taguchi model is widely used to design different parameters of different experiments or experimental models. First, the initial level is determined for each of the random parameters in the optimization algorithms. In the Taguchi method, parameters are classified into two groups: (1) controllable, and (2) uncontrollable (noise). In the Taguchi model, each parameter combination that has a higher $S$ (signal)/$N$ (noise) ratio is regarded as the best combination [58].

$$S/N = -10 \log \left( \frac{1}{n} \sum_{i=1}^{n} Y_i^2 \right) \tag{35}$$

where, $n$ is the number of data, and $Y_i$ is the fitness function that is obtained by the Taguchi model. For example, consider the PSO algorithm with four parameters and three levels. When the population size is at level 1, the acceleration coefficient is tested at levels 1, 2, 3, and 4. Similarly, the inertia coefficient is tested at levels 1, 2, 3, and 4.

*2.8. Hybrid ANN, ANFIS, and SVM Models with Optimization Algorithms*

The optimization algorithms can be used as a robust training algorithm for the ANN models. The process starts with the initialization of a group of random agents (particles, chromosomes, krill, grasshoppers, weeds, or cats). The position of agents represents the ANN weights and biases. Following this level, using the initial biases and weights (i.e., the initial position of agents), the hybrid ANN-optimization algorithms are trained, and the error between the observed and estimated value is calculated. At each iteration, the calculated error is decreased by the updating of agent locations.

The model procedure in ANFIS-optimization algorithm models starts with the initialization of a set of agents (particles, chromosomes, krill, grasshoppers, weeds, or cats) and continues with the random choice of agents and finally adjusts a location for each agent. First, the ANFIS model is trained. Then, the consequent and premise parameters are optimized by the optimization algorithms. The root mean square error (RMSE) is defined as an objective function. The aim of optimization algorithms is to minimize the objective function value with finding the appropriate values of consequent and premise parameters.

In SVM, the C parameter and kernel function parameters have significant effects on the accuracy of the SVM. The random population of agents (particles, chromosomes, krill, grasshoppers, weeds, or cats) are initialized for training the SVM parameters. The RMSE is defined as an objective function. The aim of hybrid SVM-optimization algorithm models is to minimize model errors. Figure 2 shows the developed framework of hybrid ANN, ANFIS, and SVM-optimization models for modeling groundwater level.

Thus, the model parameters are considered as decision variables for optimization algorithms. The optimization algorithms aim to minimize the error function to find the optimal value of model parameters. The PCA selects the appropriate input combinations. Then the hybrid and standalone models uses the input combinations to forecast GEWL. The models uses the optimized model parameters to accurately forecast monthly GWL.

*2.9. Uncertainty Analysis of Soft Computing Models*

The input data and the inability of model structure are the sources of uncertainty. In this research, an integrated framework is developed to simultaneously evaluate the input data and model structure.

*Input data uncertainty*

The combined Bayesian uncertainty was used to compute the uncertainty contributed by input data. The input error model was used to account for the uncertainty of input data [59]:

$$H_{a,t} = KH_t, K \sim N(m, s_m^2) \tag{366}$$

where, $H_{a,t}$: the adjusted groundwater level (GWL), $H_t$: the observed GWL, *t*: the given month, *K*: the normally distributed random, *m*: mean, and $\sigma_m$: variance. For each soft computing model, *m* and $\sigma_m$ were added to the system. A dynamically dimensioned search was used to find the value of *m*: mean and $\sigma_m$: variance as defined by [59].

**Mode Structure Uncertainty**

Bayesian model average (BMA) is used for model uncertainty. The posterior model probability and averaging over the best models were used to estimate the uncertainty of the models. The weighted average prediction of quantity of target variable is computed as follows [59]:

$$H_j = \sum_{k=1}^{k} b_k F_{jk} + e_j \tag{37}$$

where, $F_j$: the point prediction of each model, $e_j$: noise, $\beta_k$: the weight vector of model, *H*: n observation of GWL, *k*: number of models, and *j*: number of observations. For accurate application of BMA model, the standard deviation of normal probability distribution functions and weights should be estimated accurately. The log-likelihood function is used to calculate the weights and standard deviation as follows [59]:

$$L(b_{BMA}, s_{BMA} | F, H) = \sum_{i=1}^{n} \log \left\{ \sum_{k=1}^{k} b_k \frac{1}{\sqrt{2\pi s_k^2}} \exp\left[ -\frac{1}{2} s_k^{-2} (H_j - F_{jk})^2 \right] \right\} \tag{38}$$

where, $\beta_{BMA}$: maximum likelihood Bayesian weight. Markov Chain Monte Carlo (MCMC) simulations are used to compute the log-likelihood function. The integrated framework is defined as follows:

1- A number of models are selected to simulate the GWL.
2- The prior probability is assigned to each model.
3- An error input model is defined.
4- The posterior distribution of input error models and model parameters are obtained.

5- A predetermined number of GWLs for each model is provided using probabilistic parameter estimations obtained from level 2 to level 4.
6- The variance and weight of models are estimated.
7- The weights for ensemble members of models are summed to compute the weight models.
8- To the experimental soft computing models. The following indices were used to quantify the uncertainty of models:

$$p = \frac{1}{n} count[H|X_L \leq H \leq X_U] \tag{39}$$

$$d = \frac{\bar{d}_x}{\sigma_x}$$
$$\bar{d}_x = \frac{1}{k} \sum_{l=1}^{k}(X_U - X_L) \tag{40}$$

9- where $k$ is the number of observed data, $X_U$ is the upper bound of data, $X_L$ is the lower bound of data, $\sigma_x$ is standard deviation, $p$ is bracketed by 95% of predicted uncertainties, $d$ is the distance between the upper and lower bounds, and $\bar{d}_x$ is the average distance between the upper and lower bounds [59,60].

*2.10. Statistical Indices for Evaluation of Different Models*

In this study, the following indices were used to evaluate the performance of models:
Root mean square error:

$$RMSE = \sqrt{\frac{1}{N} \sum_{t=1}^{n} \left((H_0(t)) - (H_s(t))\right)^2} \tag{371}$$

Mean absolute error:

$$MAE = \frac{1}{N} \sum_{t=1}^{1} |H_0(t) - H_s(t)|^2 \tag{42}$$

Nash Sutcliffe efficiency:

$$NSE = 1 - \frac{\sum_{i=1}^{n}|H_s(t) - H_0(t)|^2}{\sum_{i=1}^{n}|H_s - \bar{H}_0(t)|^2} \tag{43}$$

Percent bias (PBIAS):

$$PBIAS = \left[\frac{\sum_{i=1}^{n}\left(H_s(t) - H_0(t)\right)^2}{\sum_{i=1}^{n}(H_o^t)^2}\right] \tag{44}$$

where, $N$ is the number of data, $H_0$ is the observed value, and $P_s$ is the predicted value.

RMSE and MAE show a good match between observed data and estimated values when it equals 0. The NSE shows a good match between the observed values and estimated values when it equals 1. The best value of PBIAS is zero.

## 3. Results and Discussion

*3.1. Inputs Selection by PCA*

In this study, 12 input variables (H(t-1), …., H(t-12)) were considered to select the input lag times of monthly GWL. As presented in the flowchart and framework of the current study in Figure 1, the first step of the model developments is the appropriate selection of time lags for GWL modelling by PCA analysis. Table 1 shows the variance contribution rate for PCAs as the principal component loadings. There are the loadings of 12 principal components versus 12 input lag times of GWL. The first four PCs variance summed up a contribution of 91%, among which the first PC variance had a contribution of 48% loadings. It was observed that the inputs H(t-1), H (t-2), H (t-3), H (t-4), and H (t-5) had higher factor loading in comparison with other inputs of the PCs. Thus, the first four PCs were selected for the hybrid soft computing models which included inputs H(t-1), H (t-2), H (t-3), H (t-4), and H (t-5) because of their higher loading factor. This loading analysis of variables reduced the raw initial input parameter numbers from 12 to 5, that decrease the model development efforts. The coefficients of more 0.75 are significant for Eigen value verifications [60].

Table 1. Principal component loadings.

| PC | 1 | 2 | 3 | 4 | 5 | 6 | 7 | 8 | 9 | 10 | 11 | 12 |
|---|---|---|---|---|---|---|---|---|---|---|---|---|
| H (t-1) | **0.98** | **0.95** | **0.93** | **0.90** | 0.89 | 0.88 | 0.88 | 0.86 | 0.75 | 0.62 | 0.52 | 0.45 |
| H (t-2) | **0.84** | **0.82** | **0.88** | **0.86** | 0.85 | 0.84 | 0.85 | 0.82 | 0.62 | 0.60 | 0.51 | 0.44 |
| H (t-3) | **0.83** | **0.81** | **0.80** | **0.77** | 0.74 | 0.72 | 0.84 | 0.80 | 0.61 | 0.55 | 0.43 | 0.40 |
| H (t-4) | **0.82** | **0.80** | **0.78** | **0.76** | 0.75 | 074 | 0.83 | 0.78 | 0.60 | 0.54 | 0.39 | 0.37 |
| H (t-5) | **0.81** | **0.79** | **0.76** | **0.75** | 0.72 | 0.71 | 0.82 | 0.79 | 0.55 | 0.51 | 0.38 | 0.35 |
| H (t-6) | 0.73 | 0.67 | 0.74 | 0.73 | 0.71 | 0.70 | 0.80 | 0.77 | 0.54 | 0.50 | 0.37 | 0.34 |
| H (t-7) | 0.62 | 0.55 | 0.72 | 0.70 | 0.65 | 0.64 | 0.76 | 0.75 | 0.53 | 0.47 | 0.33 | 0.30 |
| H (t-8) | 0.61 | 0.50 | 0.71 | 0.69 | 0.54 | 0.52 | 0.65 | 0.64 | 0.51 | 0.46 | 0.30 | 0.29 |
| H (t-9) | 0.54 | 0.54 | 0.70 | 0.65 | 0.42 | 0.64 | 0.54 | 0.52 | 0.50 | 0.45 | 0.29 | 0.25 |
| H (t-10) | 0.42 | 0.42 | 0.69 | 0.66 | 0.41 | 0.62 | 0.45 | 0.44 | 0.49 | 0.42 | 0.28 | 0.26 |
| H (t-11) | 0.42 | 0.42 | 0.55 | 0.54 | 0.40 | 0.55 | 0.42 | 0.40 | 0.47 | 0.41 | 0.27 | 0.24 |
| H (t-12) | 0.40 | 0.40 | 0.45 | 0.43 | 0.38 | 0.52 | 0.40 | 0.38 | 0.46 | 0.38 | 0.25 | 0.23 |
| Eigen value | 5.78 | 3.22 | 1.12 | 0.90 | 0.6 | 0.27 | 0.05 | 0.03 | 0.02 | 0.003 | 0.003 | 0.03 |
| Cumulative variance | 48% | 74% | 84% | 91% | 96% | 99 | 99.5 | 99.7 | 99.99 | 99.99 | 99.99 | 100% |

*3.2. Selection of Random Parameters by the Taguchi Model*

The Taguchi model was used to find the value of random parameters rather than the classical trial and error methods. Table 2 shows the computed signal-to-noise (S/N) ratio for each random parameter in the optimization module of the hybrid training of ANFIS. Each parameter had four levels and the best level of each parameter is selected based on the S/N values. The S/N ratio was computed for each level of parameters. The best value of parameters had the highest S/N rate. For example, sensitivity analysis for different values of GOA parameters was done, as shown in Table 2. The results indicated that the population size = 300 had the highest value of S/N. Thus, the optimal size of population was 300. The maximum S/N ratio for parameter l was 1.23. Thus, the optimal value of parameter l was 1.5. The maximum S/N ratio for parameter f was 1.14. Thus, the optimal value of parameter f was 0.5.

Table 2. Results of Taguchi model for a: GOA, b: particle swarm optimization (PSO), c: genetic algorithm (GA), d: WA, e: CSO, and f: krill algorithm.

| (a) | | | | | | |
|---|---|---|---|---|---|---|
| Population size | S/N | l | S/N | f | S/N |
| **100** | 1.05 | 0.5 | 1.07 | 0.1 | 1.09 |
| **200** | 1.15 | 1 | 1.19 | 0.3 | 1.12 |
| **300** | **1.20** | **1.5** | **1.23** | **0.5** | **1.14** |
| **400** | 1.02 | 2 | 1.18 | 0.7 | 1.10 |
| (b) | | | | | | |
| Population size | S/N | $c_1$ | S/N | $c_2$ | S/N | $\omega$ | S/N |
| **100** | 1.25 | 1.6 | 1.20 | 1.6 | 1.21 | 0.3 | 1.19 |
| **200** | 1.29 | 1.8 | 1.27 | 1.8 | 1.25 | 0.50 | 1.18 |
| **300** | 1.23 | 2.0 | 1.26 | 2.0 | 1.23 | 0.70 | 1.17 |
| **400** | 1.20 | 2.2 | 1.22 | 2.2 | 1.25 | 0.90 | 1.24 |

(c)

| **Population size** | **S/N** | **Mutation probability** | **S/N** | **Crossover rate** | **S/N** |
|---|---|---|---|---|---|
| **100** | 1.18 | 0.01 | 1.16 | 1.6 | 1.21 |
| **200** | 1.20 | 0.03 | 1.17 | 1.8 | 1.25 |
| **300** | 1.21 | 0.05 | 1.20 | 2.0 | 1.23 |
| **400** | 1.17 | 0.07 | 1.19 | 2.2 | 1.25 |

(d)

| **$P_{max}$** | **S/N** | **n** | **S/N** |
|---|---|---|---|
| **50** | 1.12 | 1 | 1.14 |
| **100** | 1.23 | 2 | 1.17 |
| **150** | 1.19 | 3 | 1.18 |
| **200** | 1.17 | 4 | 1.19 |

(e)

| **Population size** | **S/N** | **SMP** | **S/N** | **MR** | **S/N** |
|---|---|---|---|---|---|
| 100 | 1.11 | 5 | 1.10 | 0.10 | 1.12 |
| 200 | 1.24 | 10 | 1.15 | 0.30 | 1.16 |
| 300 | 1.17 | 15 | 1.17 | 0.50 | 1.18 |
| 400 | 1.15 | 20 | 1.21 | 0.70 | 1.20 |

(f)

| **Population size** | **S/N** | **$V_f$** | **S/N** | **$N_{max}$** | **S/N** |
|---|---|---|---|---|---|
| 100 | 1.10 | 0.005 | 1.12 | 0.02 | 1.14 |
| 200 | 1.12 | 0.010 | 1.15 | 0.04 | 1.17 |
| 300 | 1.14 | 0.015 | 1.17 | 0.06 | 1.12 |
| 400 | 1.16 | 0.020 | 1.14 | 0.08 | 1.21 |

*3.3. Results of Hybrid ANN, ANFIS, and SVM Models*

In this section, the results of developed hybrid models are presented and compared with each other and with the usual ANFIS, ANN, and SVM models. These models are hybridized with GOA, CSO, KA, WA, PSO, and GA meta-heuristic optimization algorithms. The results of models in three piezometers of 6, 9, and 10 as shown in Figure 6, are presented and discussed. These piezometers were selected as samples to evaluate the ability of new hybrid models.

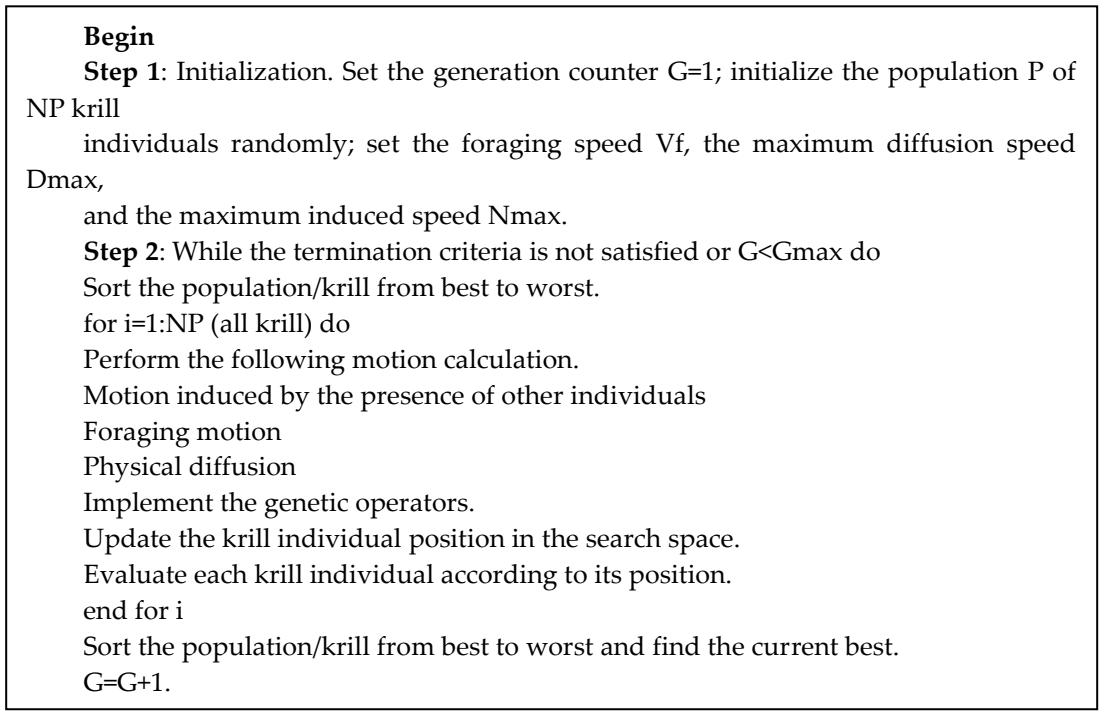

```
Begin
Step 1: Initialization. Set the generation counter G=1; initialize the population P of NP krill
    individuals randomly; set the foraging speed Vf, the maximum diffusion speed Dmax,
    and the maximum induced speed Nmax.
Step 2: While the termination criteria is not satisfied or G<Gmax do
Sort the population/krill from best to worst.
for i=1:NP (all krill) do
Perform the following motion calculation.
Motion induced by the presence of other individuals
Foraging motion
Physical diffusion
Implement the genetic operators.
Update the krill individual position in the search space.
Evaluate each krill individual according to its position.
end for i
Sort the population/krill from best to worst and find the current best.
G=G+1.
```

**Figure 6.** The flowchart of the krill algorithm [54].

- piezometer 6

Table 3 and Figure 7a show the results of hybrid optimized and standalone soft computing models for piezometer 6. Results indicated that ANFIS-GOA was the most accurate model and is selected as the optimum model that was verified by a value of RMSE = 1.12 m, MAE = 0.812 m, NSE = 0.95, and PBIAS = 0.12 for the training level. For the testing phase assessed with the ANFIS-GOA, results indicated a value of RMSE: 1.21 m, MAE: 0.878 m, NSE: 0.93, and PBIAS: 0.15 which reflected better performance in comparison to other models. From Table 3, results indicated that the SVM model with the higher values of RMSE, MAE, and PBIAS and lower values of NSE was the worst model among other models. Among the hybrid ANN models, the ANN-GOA outperformed the ANN-CSO, ANN-GA, ANN-PSO, ANN-WA, and ANN-KA models with the best values for RMSE = 1.21 m, MAE = 0.878 m, NSE = 0.93, PBIAS = 0.15 in the test stage. The ability of GA was lower than that of CSO, PSO, WA, and KA because of higher values of RMSE, MAE, and PBIAS and lower values of NSE in train and test steps as presented in Table 3. Among SVM models, the hybrid SVM-GOA was observed to have the lowest value of NSE and the highest values of RMSE, MAE, and PBIAS. It was important to mention that the standalone SVM, ANN, and ANFIS had worse performance than hybrid ANN, SVM, and ANFIS models that indicates the superiority of hybridization in model developments. Among PSO, CSO, GA, KA, and WA, the CSO had better results than the other optimization algorithms. The general results showed that ANFSI model was superior to the SVM and ANN models. Additionally, the ANN model had lower values of RMSE and MAE than did the SVM model. Additionally, the results of ANFIS-GOA as the best model in piezometer 6 in comparison with standalone ANFIS shows that meta-heuristic hybridizations improved the model performances in train and test steps. The percent of RMSE, MAE, NSE, and PBIAS improvements by ANFIS-GOA in train step were 15%, 4%, 13%, and 208% and these values for the test steps of ANFIS-GOA are 33%, 44.6%, 16.3%, and 173%, respectively, that clearly confirm the superiority of developed hybridization schemes in GWL modelling. Additionally, in Figure 7a, the scatter plots of training and testing steps visualize the performance of ANFIS-GOA compared to the other models. Furthermore, simulations coincide very well with the observed values and all of the data points concentrated over the y = x line with $R^2$ = 0.93. Furthermore, this figure shows that other hybridized models such as CSO, PSO, KA, WA, GA, and standalone ANFIS have less accuracy in high and low values of GWL, while the ANFIS-GOA over all of low to high values of GWL performed accurately in regard to the observations.

Table 3. Statistical characteristics of applied hybrid models for piezometer 6.

| Model | Training | | | | | Testing | | | | |
|---|---|---|---|---|---|---|---|---|---|---|
| | RMSE | MAE | NSE | PBIAS | $R^2$ | RMSE | MAE | NSE | PBIAS | $R^2$ |
| ANFIS-GOA | **1.12** | **0.812** | **0.95** | **0.12** | **0.95** | **1.21** | **0.878** | **0.93** | **0.15** | **0.93** |
| ANN-GOA | 1.24 | 0.815 | 0.92 | 0.14 | **0.94** | 1.25 | 0.897 | 0.91 | 0.16 | 0.92 |
| SVM-GOA | 1.25 | 0.817 | 0.91 | 0.17 | **0.91** | 1.29 | 0.901 | 0.90 | 0.18 | 0.90 |
| ANFIS-CSO | 1.14 | 0.819 | 0.94 | 0.15 | **0.94** | 1.30 | 0.899 | 0.92 | 0.17 | 0.92 |
| ANN-CSO | 1.28 | 0.821 | 0.93 | 0.18 | **0.93** | 1.34 | 0.935 | 0.90 | 0.19 | 0.90 |
| SVM-CSO | 1.32 | 0.823 | 0.90 | 0.20 | **0.89** | 1.38 | 0.939 | 0.89 | 0.22 | 0.87 |
| ANFIS-KA | 1.19 | 0.825 | 0.93 | 0.16 | **0.93** | 1.41 | 1.01 | 0.91 | 0.24 | 0.90 |
| ANN-KA | 1.30 | 0.829 | 0.91 | 0.22 | **0.90** | 1.42 | 1.09 | 0.89 | 0.25 | 0.88 |
| SVM-KA | 1.33 | 0.832 | 0.89 | 0.24 | **0.88** | 1.43 | 1.12 | 0.87 | 0.26 | 0.85 |
| ANFIS-WA | 1.21 | 0.827 | 0.92 | 0.27 | **0.92** | 1.45 | 1.10 | 0.89 | 0.28 | 0.93 |
| ANN-WA | 1.32 | 0.832 | 0.90 | 0.29 | **0.90** | 1.47 | 1.14 | 0.86 | 0.31 | 0.87 |
| SVM-WA | 1.35 | 0.833 | 0.88 | 0.33 | **0.84** | 1.51 | 1.16 | 0.85 | 0.35 | 0.83 |
| ANFIS-PSO | 1.24 | 0.829 | 0.88 | 0.35 | **0.90** | 1.53 | 1.12 | 0.84 | 0.37 | 0.89 |
| ANN-PSO | 1.35 | 0.835 | 0.87 | 0.37 | **0.89** | 1.55 | 1.17 | 0.85 | 0.39 | 0.86 |
| SVM-PSO | 1.37 | 0.839 | 0.86 | 0.39 | **0.83** | 1.52 | 1.19 | 0.83 | 0.43 | 0.82 |
| ANFIS-GA | 1.28 | 0.835 | 0.87 | 0.35 | **0.88** | 1.59 | 1.21 | 0.82 | 0.37 | 0.87 |
| ANN-GA | 1.32 | 0.839 | 0.85 | 0.39 | **0.87** | 1.62 | 1.23 | 0.81 | 0.40 | 0.84 |
| SVM-GA | 1.35 | 0.842 | 0.83 | 0.41 | **0.82** | 1.71 | 1.25 | 0.80 | 0.42 | 0.81 |
| ANFIS | 1.30 | 0.844 | 0.84 | 0.37 | **0.85** | 1.61 | 1.27 | 0.80 | 0.41 | 0.83 |
| ANN | 1.38 | 0.849 | 0.82 | 0.43 | **0.87** | 1.73 | 1.29 | 0.78 | 0.45 | 0.84 |
| SVM | 1.40 | 0.851 | 0.81 | 0.45 | **0.80** | 1.75 | 1.32 | 0.77 | 0.47 | 0.79 |

- piezometer 9

Results of hybrid models for piezometer 9 in Figure 7b and Table 4 indicated that the hybrid ANN, ANFIS, and SVM models had better performance than the standalone ANN, SVM, and ANFIS models, the same as the results for piezometer 6 in the previous subsection. Among ANFIS hybrid models, the hybrid ANFIS-GOA was confirmed to have the best performance with the smallest values of RMSE = 1.16 m, MAE = 0.818 m, and PBIAS = 0.14 and the highest values of NSE = 0.94 in the training stage and in testing stage these values were 1.22 m, 0.881 m, 0.17, and 0.92 respectively. The ANFIS model provided the best RMSE, PBIAS, MAE, and NSE among other models. The best values of RMSE, MAE, PBIAS, and NSE for ANN-GOA in the training phase were 1.25 m, 0.819 m, 0.91, and 0.19, respectively. Results indicated that the SVM model had the worst performance among other models. For the testing phase assessed with SVM-GOA, the results indicated a value of RMSE: 1.31 m, MAE: 0.903 m, NSE: 0.89, and PBIAS: 0.20 which reflected better performance than the SVM model and indicates the improvements when SVM is hybridized with the GOA. Results of Table 4 indicated that GOA and GA were the best and worst algorithms among other algorithms. As observed in Table 4 and Figure 7b, the evolutionary ANN models had more accuracy than the evolutionary SVM model because of lower values of RMSE, MAE, and PBIAS and higher values of NSE. However, no major differences were observed in GWL predictions of piezometers 6 and 9 predictions by ANFIS-GOA.

Table 4. Statistical characteristics of applied hybrid models for piezometer 9.

| Model | Training | | | | | Testing | | | | |
|---|---|---|---|---|---|---|---|---|---|---|
| | RMSE | MAE | NSE | PBIAS | $R^2$ | RMSE | MAE | NSE | PBIAS | $R^2$ |
| ANFIS-GOA | **1.16** | **0.818** | **0.94** | **0.14** | **0.96** | **1.22** | **0.881** | **0.92** | **0.17** | **0.94** |
| ANN-GOA | 1.25 | 0.819 | 0.91 | 0.15 | **0.95** | 1.27 | 0.899 | 0.90 | 0.18 | 0.93 |
| SVM-GOA | 1.27 | 0.821 | 0.90 | 0.19 | **0.91** | 1.31 | 0.903 | 0.89 | 0.20 | 0.90 |
| ANFIS-CSO | 1.18 | 0.820 | 0.93 | 0.16 | **0.95** | 1.32 | 0.901 | 0.91 | 0.19 | 0.93 |

| Model | | | | | | | | | | |
|---|---|---|---|---|---|---|---|---|---|---|
| ANN-CSO | 1.29 | 0.823 | 0.92 | 0.19 | **0.94** | 1.36 | 0.938 | 0.88 | 0.18 | 0.92 |
| SVM-CSO | 1.33 | 0.825 | 0.91 | 0.22 | **0.89** | 1.39 | 0.940 | 0.87 | 0.20 | 0.88 |
| ANFIS-KA | 1.20 | 0.827 | 0.92 | 0.18 | **0.94** | 1.34 | 1.05 | 0.90 | 0.22 | 0.92 |
| ANN-KA | 1.31 | 0.831 | 0.90 | 0.23 | **0.91** | 1.44 | 1.10 | 0.86 | 0.23 | 0.90 |
| SVM-KA | 1.35 | 0.833 | 0.88 | 0.25 | **0.87** | 1.45 | 1.14 | 0.85 | 0.27 | 0.86 |
| ANFIS-WA | 1.22 | 0.829 | 0.91 | 0.28 | **0.91** | 1.49 | 1.12 | 0.83 | 0.29 | 0.90 |
| ANN-WA | 1.36 | 0.834 | 0.89 | 0.30 | **0.89** | 1.51 | 1.15 | 0.82 | 0.32 | 0.88 |
| SVM-WA | 1.38 | 0.835 | 0.87 | 0.34 | **0.86** | 1.53 | 1.17 | 0.83 | 0.37 | 0.85 |
| ANFIS-PSO | 1.27 | 0.831 | 0.86 | 0.36 | **0.87** | 1.55 | 1.19 | 0.81 | 0.39 | 0.86 |
| ANN-PSO | 1.39 | 0.837 | 0.85 | 0.38 | **0.87** | 1.57 | 1.23 | 0.80 | 0.40 | 0.85 |
| SVM-PSO | 1.40 | 0.840 | 0.84 | 0.40 | **0.85** | 1.59 | 1.25 | 0.83 | 0.45 | 0.84 |
| ANFIS-GA | 1.29 | 0.839 | 0.83 | 0.39 | **0.85** | 1.61 | 1.28 | 0.80 | 0.39 | 0.84 |
| ANN-GA | 1.42 | 0.840 | 0.82 | 0.40 | **0.86** | 1.63 | 1.29 | 0.79 | 0.42 | 0.84 |
| SVM-GA | 1.43 | 0.843 | 0.81 | 0.42 | **0.82** | 1.69 | 1.32 | 0.77 | 0.43 | 0.81 |
| ANFIS | 1.33 | 0.845 | 0.82 | 0.39 | **0.84** | 1.71 | 1.39 | 0.79 | 0.42 | 0.83 |
| ANN | 1.44 | 0.851 | 0.80 | 0.44 | **0.85** | 1.76 | 1.40 | 0.77 | 0.47 | 0.83 |
| SVM | 1.45 | 0.852 | 0.79 | 0.47 | **0.81** | 1.77 | 1.43 | 0.76 | 0.49 | 0.78 |

- piezometer 10

Here the results of models in piezometer 10 are evaluated. As observed in Table 5, results indicated that the ANFIS-GOA was better in terms of minimizing RMSE, MAE, and PBIAS than the other models. ANFIS-GOA reduced RMSE error by 7.01% and 7.04% compared to ANN-GOA and SVM-GOA, respectively. The standalone ANFIS, ANN, and SVM models provided worse results than the hybrid models. The SVM model provided the worst performance among other models. The NSE of ANFIS-GOA, ANFIS-CSO, ANFIS-KA, ANFIS-WA, ANFIS-PSO, and ANFIS-GA was 0.91, 0.90, 0.89, 0.79, and 0.75, respectively. GA had the worst performance among other algorithms. As is shown in Table 5, the error in the estimated GWL by using GA was more than that of PSO, KA, WA, GA, CSO, and GOA. Overall, the percent of improvements in the ANFIS-GOA versus standalone ANFIS in piezometer 6 were 14.4%, 3%, 17.8%, and 181% for RMSE, MAE, NSE, and PBIAS in training stage and 40.7%, 55%, 25%, and 132% in testing stage, respectively. These values again confirm that all of the hybridized models performed more accurately than the stand-alone models and indicate the generality of hybridizing Taguchi with training procedure compared to the classical standalone models.

Table 5. Statistical characteristics of applied hybrid models for piezometer 10.

| Model | Training | | | | | Testing | | | | |
|---|---|---|---|---|---|---|---|---|---|---|
| | RMSE | MAE | NSE | PBIAS | $R^2$ | RMSE | MAE | NSE | PBIAS | $R^2$ |
| ANFIS-GOA | **1.18** | **0.819** | **0.93** | **0.16** | **0.96** | 1.23 | 0.911 | 0.91 | 0.19 | 0.94 |
| ANN-GOA | 1.27 | 0.821 | 0.90 | 0.17 | **0.95** | 1.28 | 0.921 | 0.90 | 0.20 | 0.94 |
| SVM-GOA | 1.29 | 0.823 | 0.89 | 0.20 | **0.89** | 1.32 | 0.925 | 0.87 | 0.21 | 0.88 |
| ANFIS-CSO | 1.20 | 0.822 | 0.92 | 0.17 | **0.95** | 1.34 | 0.914 | 0.90 | 0.22 | 0.93 |
| ANN-CSO | 1.31 | 0.824 | 0.91 | 0.20 | **0.92** | 1.37 | 0.926 | 0.87 | 0.23 | 0.90 |
| SVM-CSO | 1.35 | 0.827 | 0.90 | 0.23 | **0.87** | 1.40 | 0.930 | 0.86 | 0.25 | 0.86 |
| ANFIS-KA | 1.22 | 0.829 | 0.89 | 0.19 | **0.94** | 1.41 | 1.10 | 0.89 | 0.24 | 0.91 |
| ANN-KA | 1.33 | 0.833 | 0.87 | 0.24 | **0.89** | 1.43 | 1.12 | 0.85 | 0.26 | 0.88 |
| SVM-KA | 1.37 | 0.835 | 0.86 | 0.27 | **0.85** | 1.47 | 1.17 | 0.84 | 0.28 | 0.84 |
| ANFIS-WA | 1.24 | 0.837 | 0.90 | 0.29 | **0.90** | 1.50 | 1.14 | 0.82 | 0.30 | 0.89 |
| ANN-WA | 1.37 | 0.839 | 0.88 | 0.31 | **0.88** | 1.52 | 1.16 | 0.81 | 0.33 | 0.87 |
| SVM-WA | 1.39 | 0.840 | 0.86 | 0.35 | **0.84** | 1.54 | 1.18 | 0.80 | 0.38 | 0.82 |
| ANFIS-PSO | 1.29 | 0.838 | 0.85 | 0.37 | **0.89** | 1.56 | 1.20 | 0.79 | 0.40 | 0.87 |
| ANN-PSO | 1.40 | 0.842 | 0.84 | 0.39 | **0.87** | 1.58 | 1.25 | 0.78 | 0.41 | 0.86 |
| SVM-PSO | 1.41 | 0.844 | 0.83 | 0.41 | **0.83** | 1.60 | 1.27 | 0.77 | 0.43 | 0.81 |

| | | | | | | | | | |
|---|---|---|---|---|---|---|---|---|---|
| ANFIS-GA | 1.31 | 0.839 | 0.82 | 0.42 | **0.86** | 1.62 | 1.29 | 0.76 | 0.42 | 0.85 |
| ANN-GA | 1.44 | 0.845 | 0.81 | 0.43 | **0.88** | 1.65 | 1.32 | 0.75 | 0.44 | 0.85 |
| SVM-GA | 1.45 | 0.847 | 0.80 | 0.44 | **0.82** | 1.71 | 1.33 | 0.74 | 0.45 | 0.80 |
| ANFIS | 1.35 | 0.849 | 0.79 | 0.45 | **0.85** | 1.73 | 1.41 | 0.73 | 0.44 | 0.84 |
| ANN | 1.45 | 0.853 | 0.78 | 0.47 | **0.85** | 1.77 | 1.42 | 0.72 | 0.49 | 0.82 |
| SVM | 1.47 | 0.855 | 0.77 | 0.49 | **0.8** | 1.78 | 1.45 | 0.70 | 0.50 | 0.79 |

*3.4. Analysis of Scatterplots of Soft Computing Models*

- piezometer 6

Scatterplots for the soft computing models are provided in Figure 7a for the training and testing phases. It is clear that the hybrid ANFIS-GOA predictions were much closer to the measured data in the testing and training phases with a higher coefficient of determination. This result indicated a better correlation and a larger degree of statistical match between measured and predicted data of ANFIS-GOA relative to the other hybrid ANN and SVM models. The $R^2$ values were found to vary in the range of 0.84–0.94 and 0.79–0.91 for the ANN (hybrid ANN models and based ANN model) and SVM models (hybrid SVM models and based SVM model), respectively. The SVM model had the lowest $R^2$ among other models. Additionally, the ANFIS-GA, ANN-GA, and SVM-GA models had the lowest $R^2$ among other hybrid ANFIS, ANN, and SVM models. There is a weak agreement between the lower and higher values of the actual and estimated GWLs in this scatter plots of piezometer 6, unlike the ANFIS-GOA results.

- piezometer 9

As observed in Figure 7b, the $R^2$ values of testing phase were 0.94, 0.93, 0.92, 0.90, 0.86, 0.84, and 0.83 for ANFIS-GOA, ANFIS-CSO, ANFIS-KA, ANFIS-WA, ANFIS-PSO, ANFIS-GA, and ANFIS model, respectively. GOA had a better performance than other optimization algorithms. The outputs indicated that all hybrid optimized ANFIS, ANN, and SVM models outperformed the standalone ANFIS, ANN, and SVM models. As the results in Table 4 show incorporating the Taguchi and GOA in ANFIS training enhanced the $R^2$ values 13% in comparison with the standalone ANFIS and in all of the developed models the hybridized meta-heuristic models outperformed the single standalone models.

- Piezometer 10

The results of Figure 7c indicated that the ANFIS-GOA and SVM models produced the best and the worst results, respectively. It is clear that developed hybrid ANFIS-GOA model forecasting of GWL was less scattered and closer to the straight line of 1:1 than those the other models and it shows impressive results in regard to the other models. For training and testing phases, GA had a worse performance than CSO, PSO, KA, WA, and GOA because of the lower values of $R^2$. The standalone ANFIS model had the worst performance among the ANFIS-GOA, ANFIS-CSO, ANFIS-WA, ANFIS-PSO, ANFIS-GA, and ANFIS-KA models. The ANFIS-GOA model with $R^2 = 0.94$ as is presented in Table 5, the values of GWL simulated by the ANFIS-GOA are almost equal to the observed values of GWL. The linear fit of the forecasted GWL and measured GWL results have a high correlation coefficient that is very close to 1.00 ($R^2 = 0.97$) and a perfect correlation coefficient ($R^2$ value) of 0.94, confirmed that the simulation model has provided a very good prediction of the observed values of GWL. Additionally, 94% of the observed GWL values accurately fit the hybrid ANFIS-GOA model predictions.

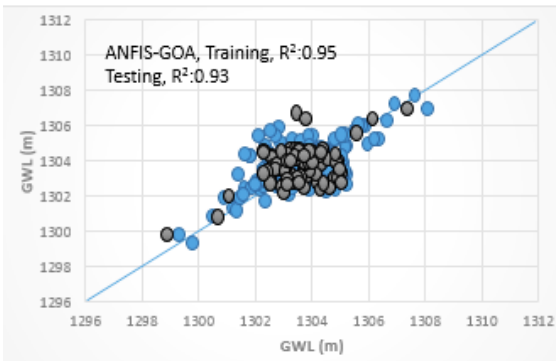
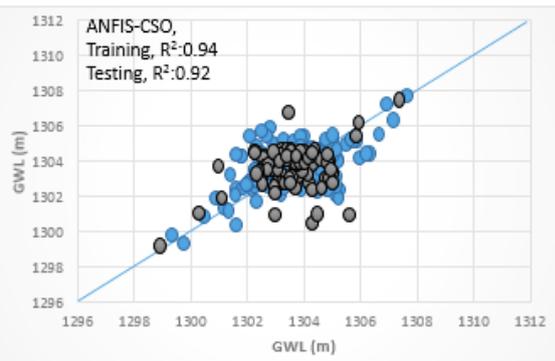
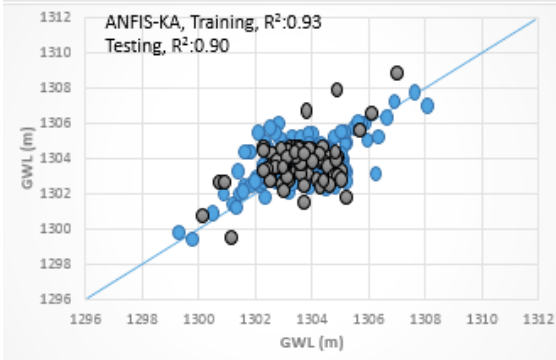
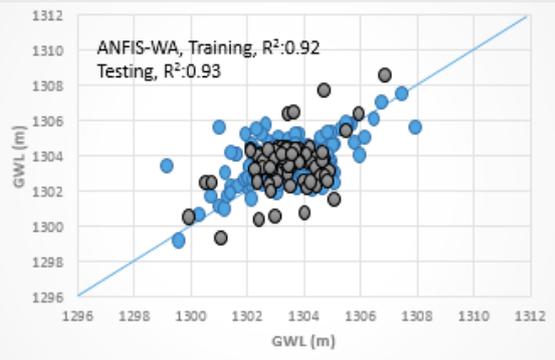
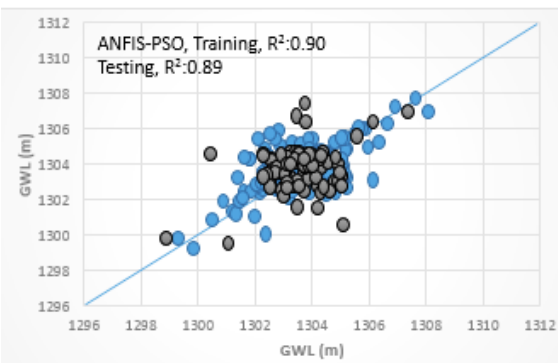
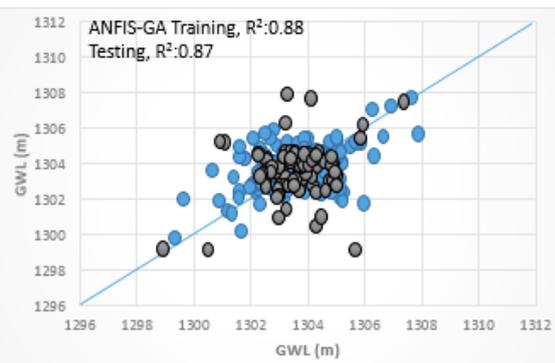
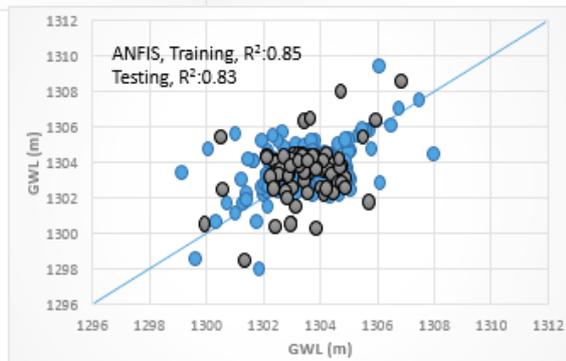

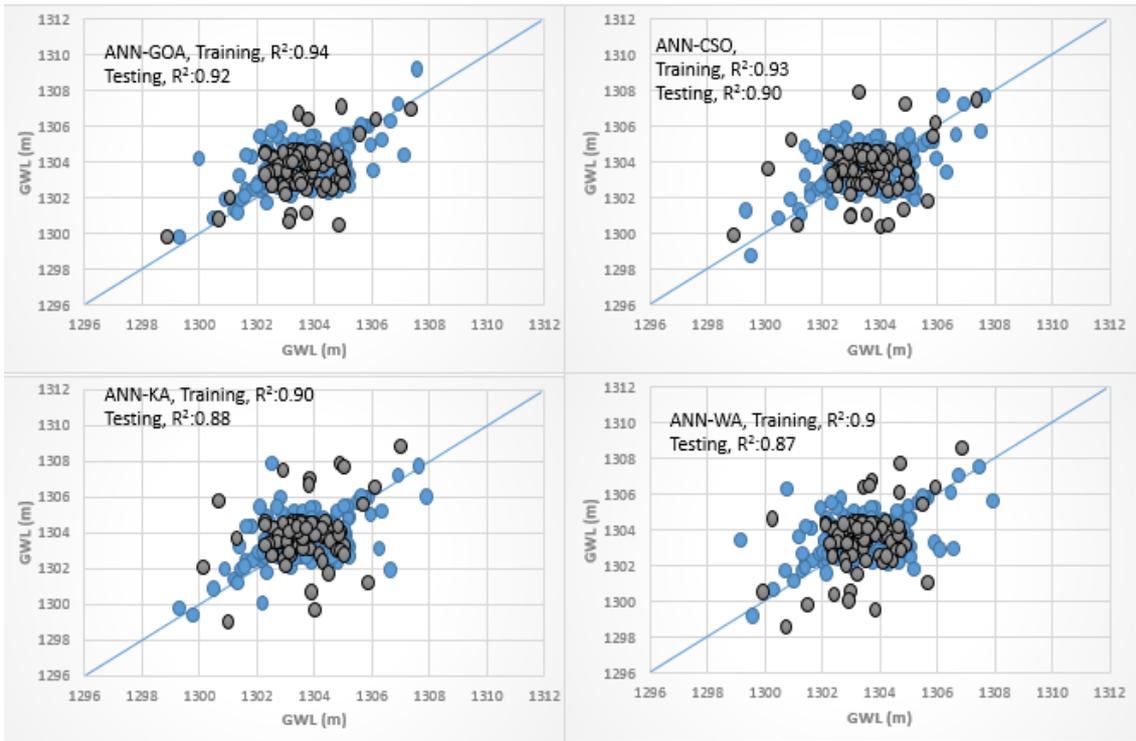
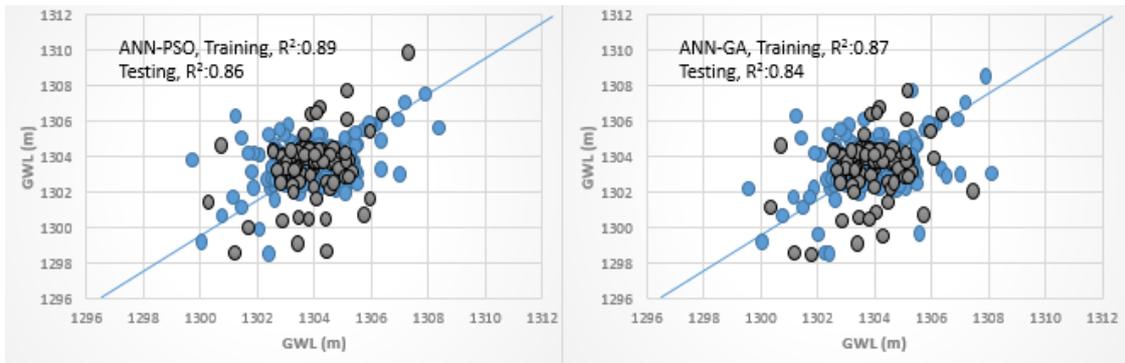
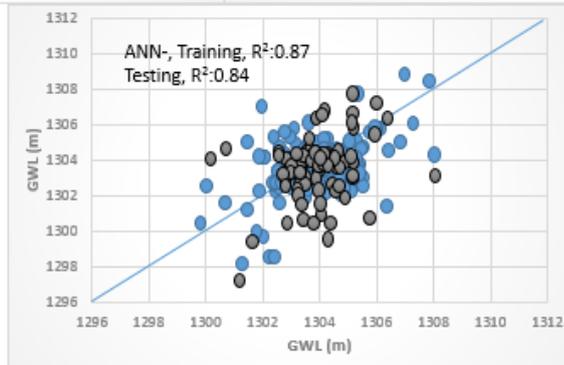

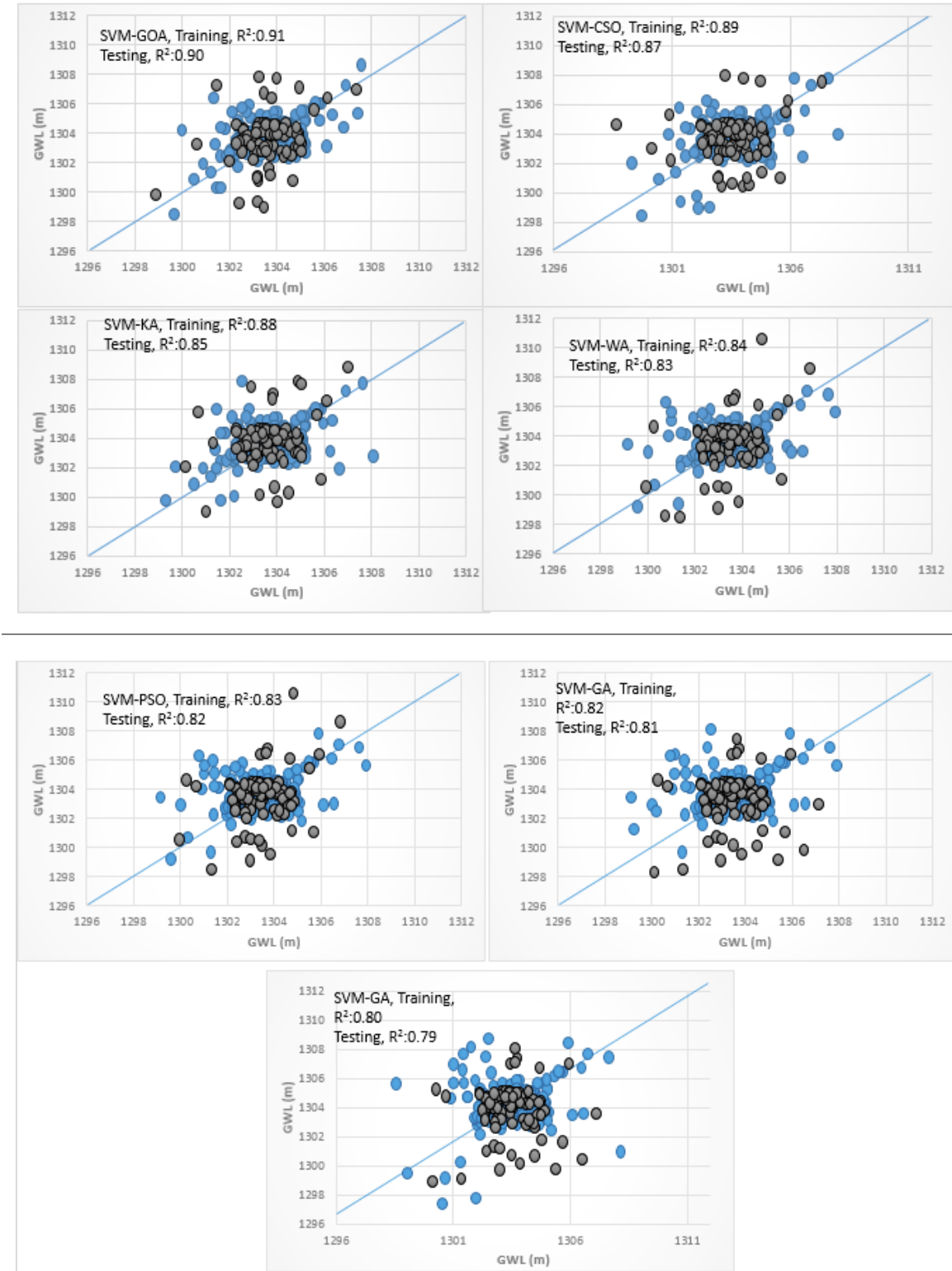

(a)

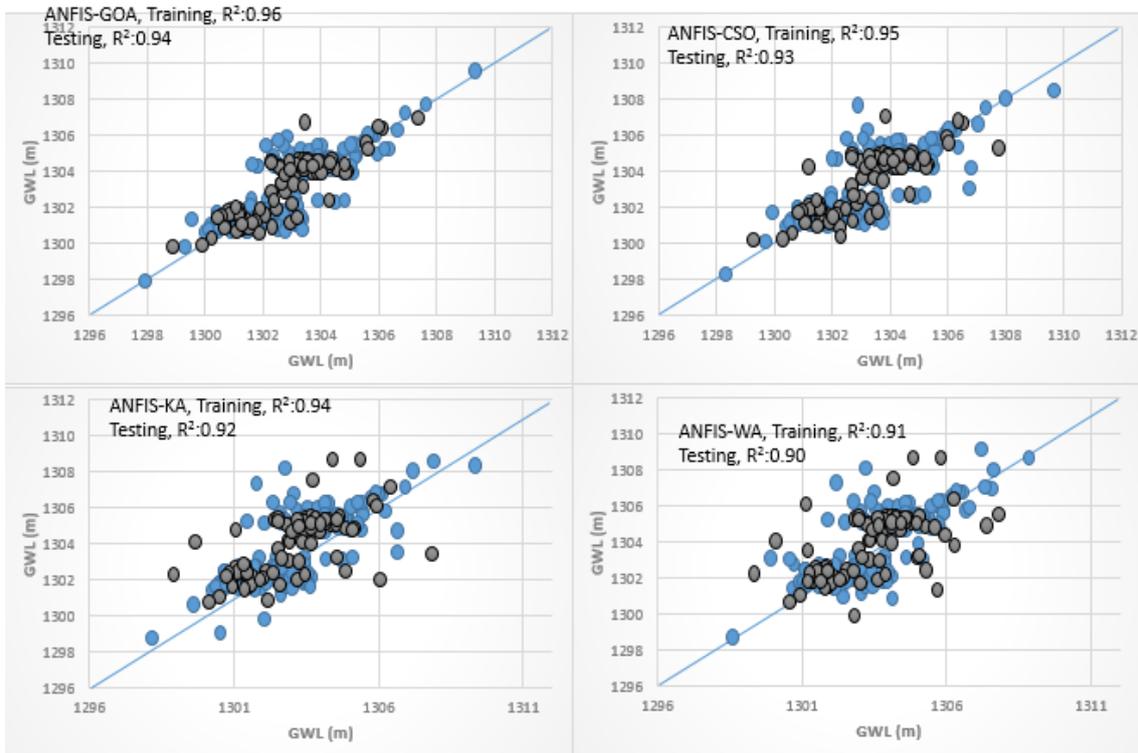

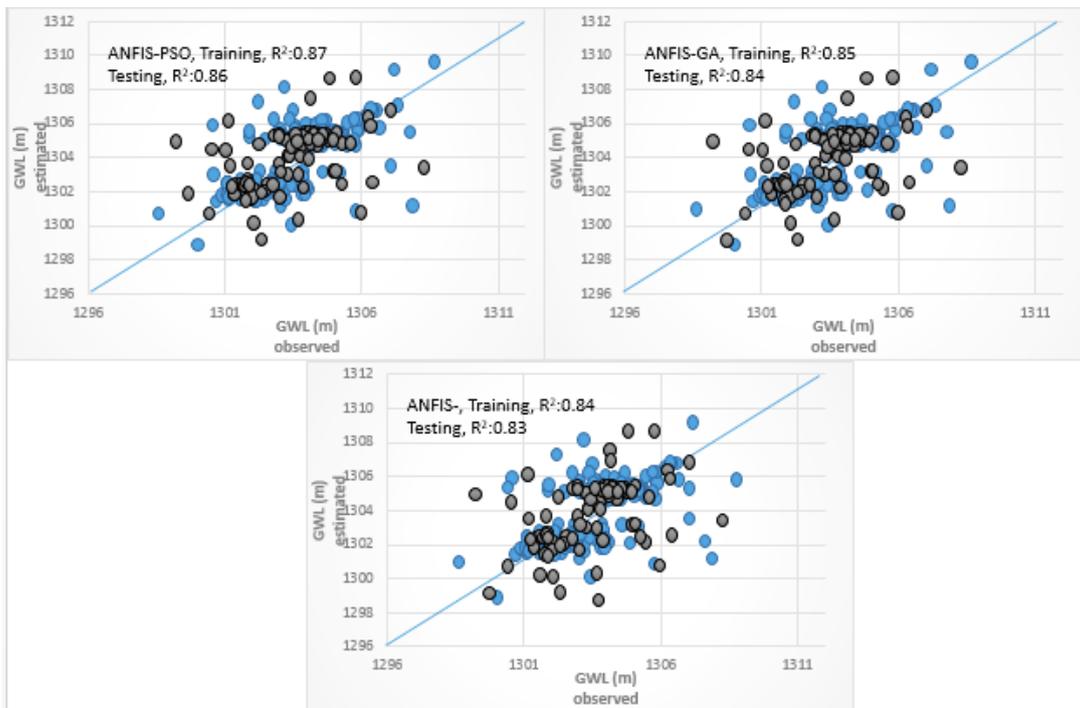

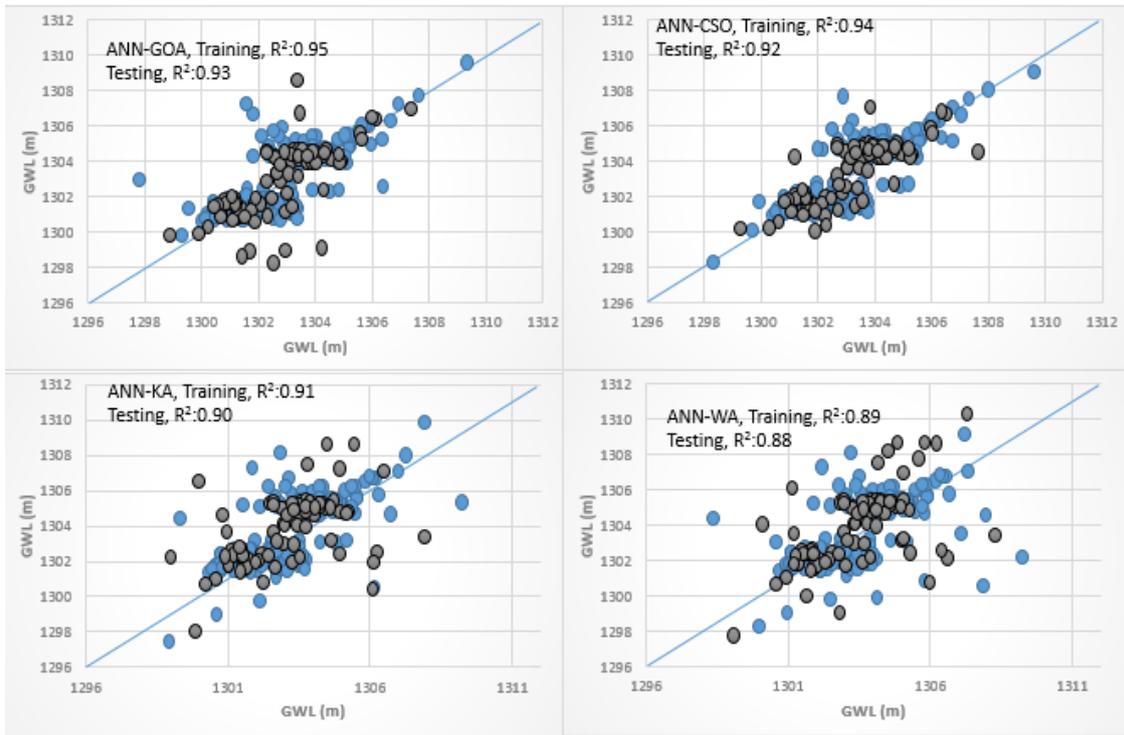
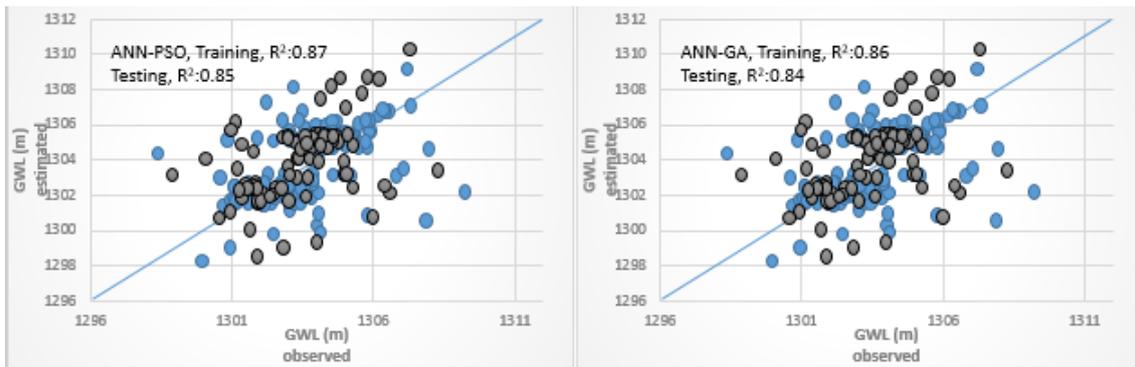
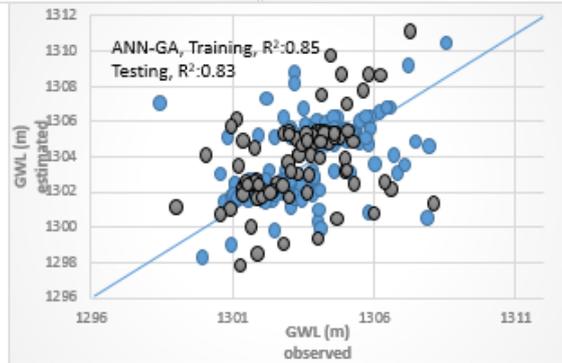

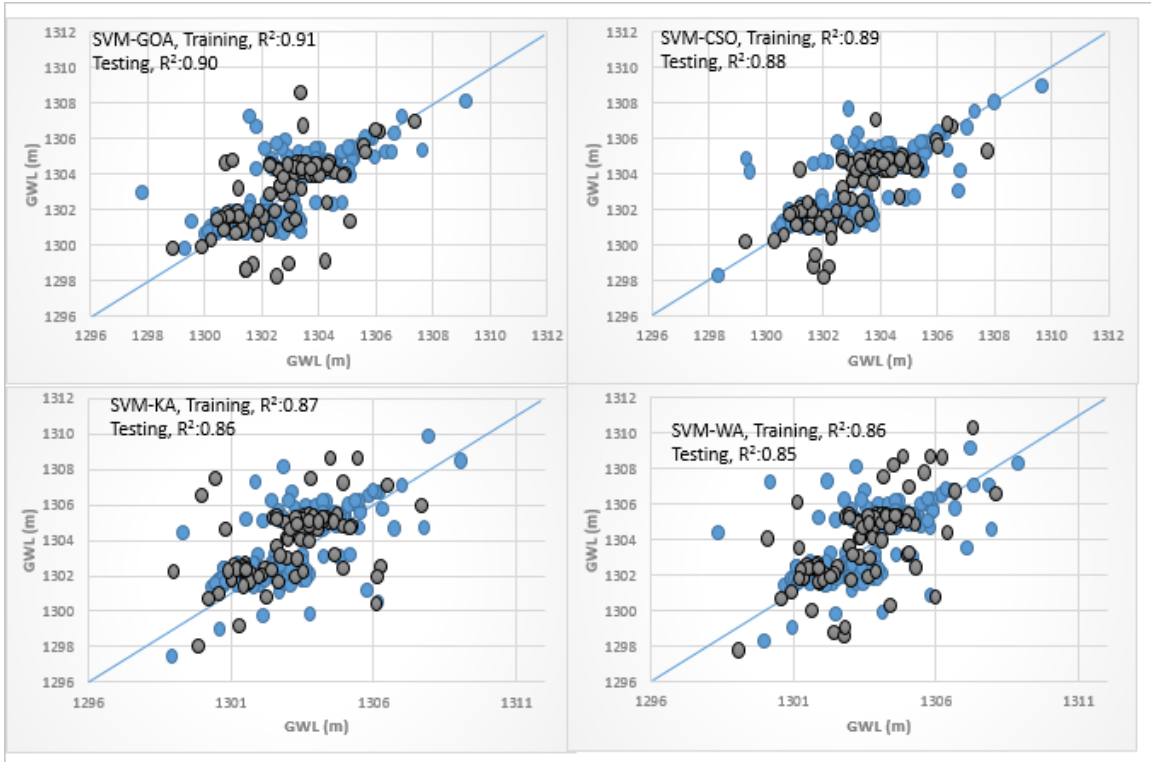
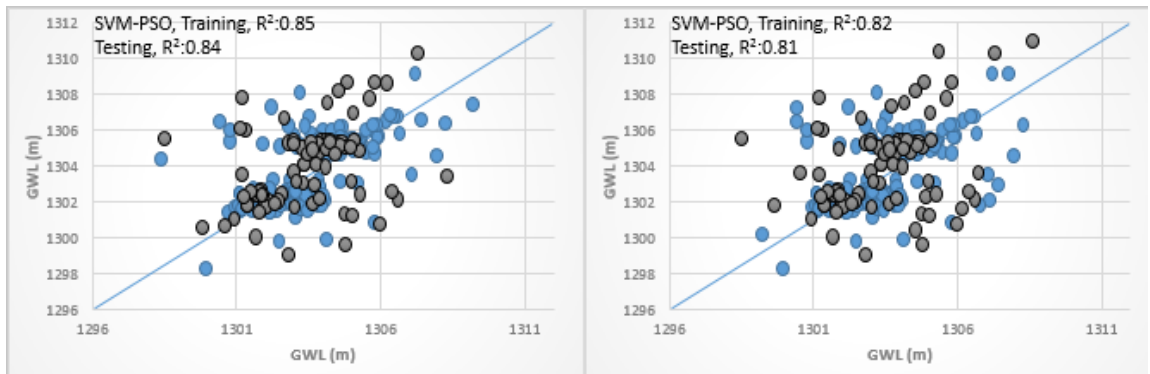
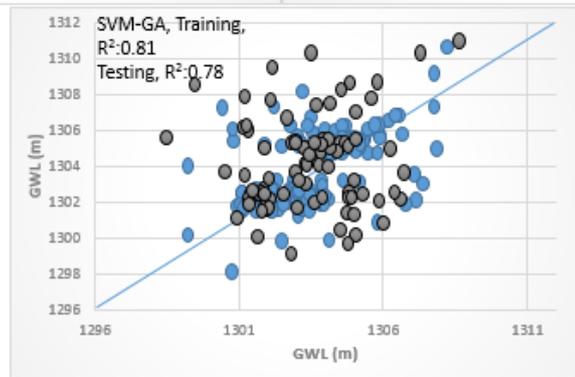

**(b)**

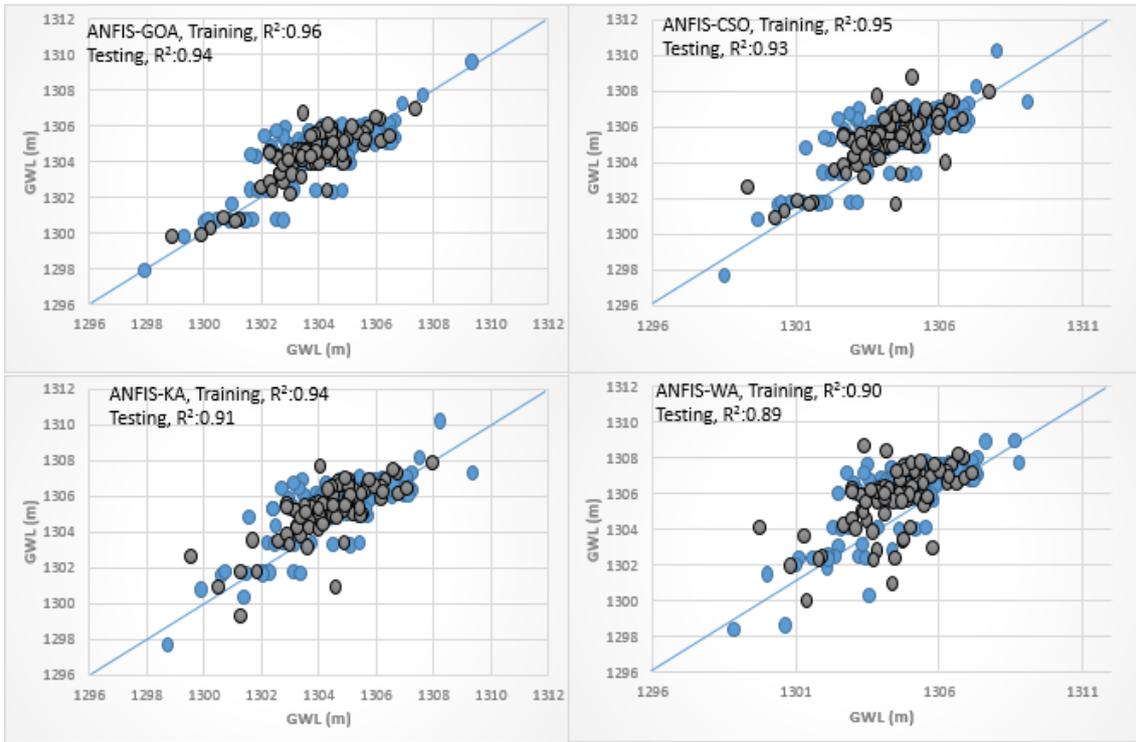
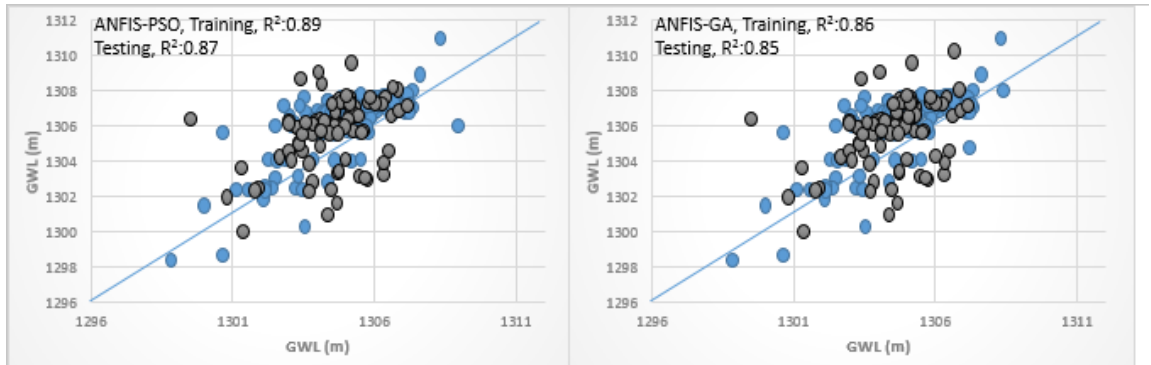
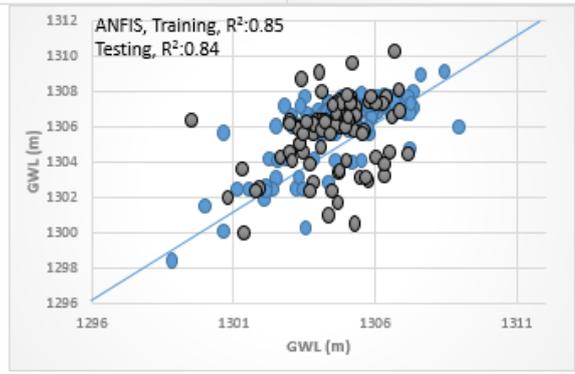

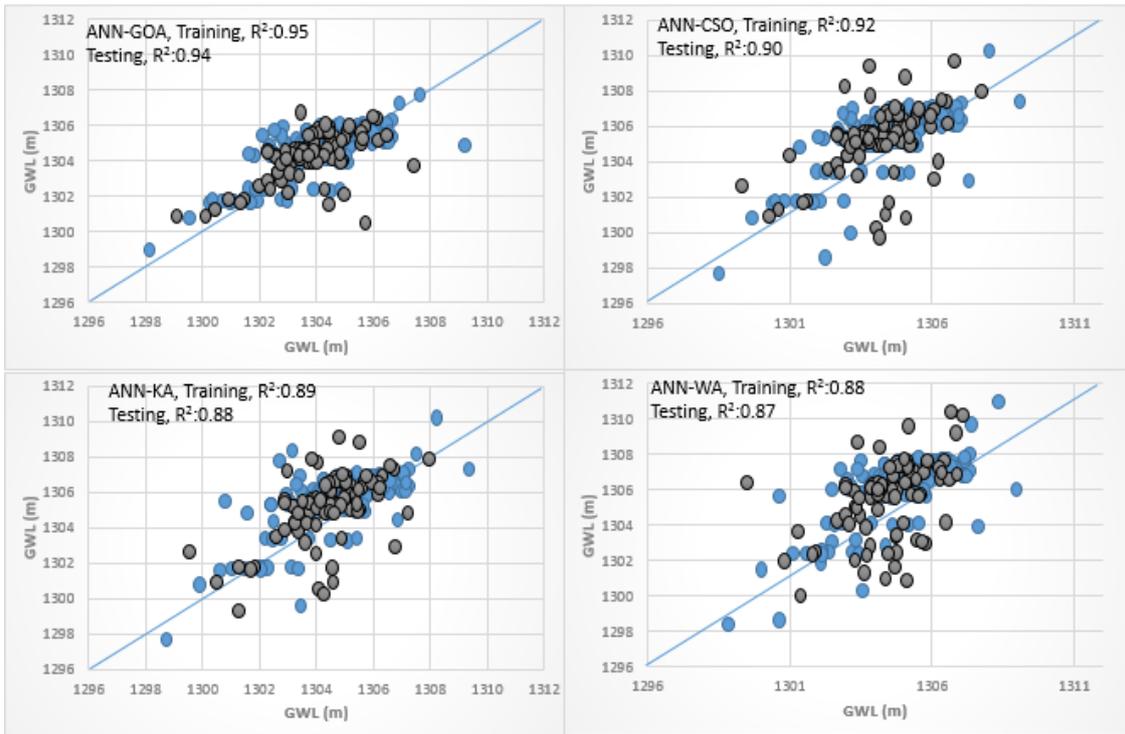
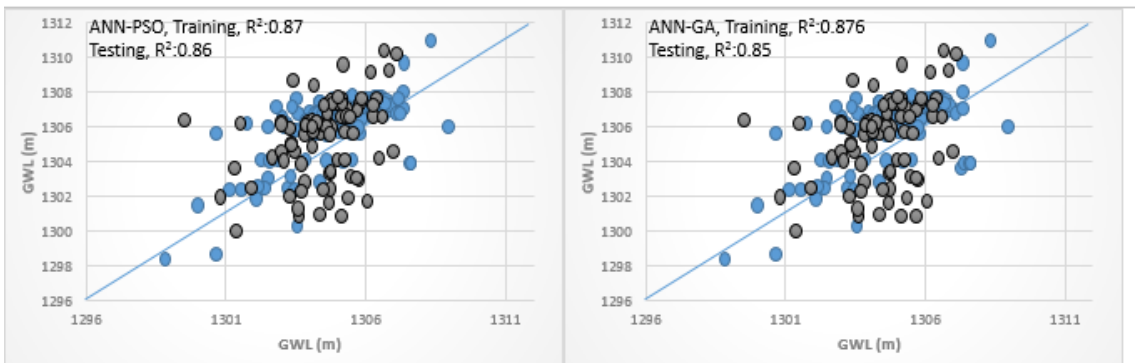
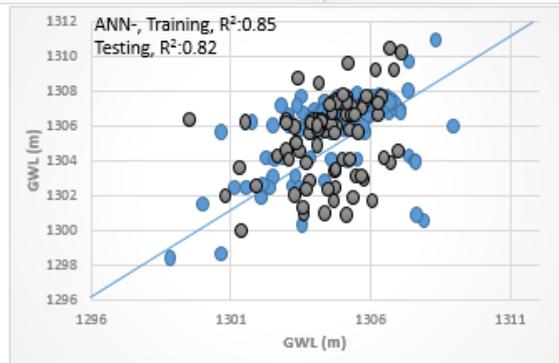

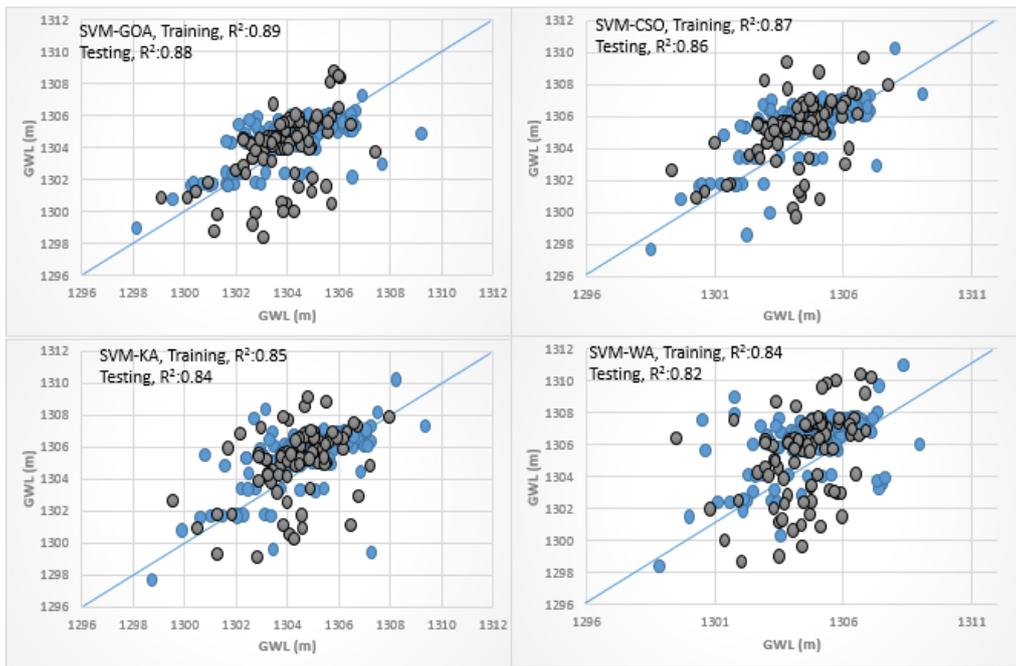

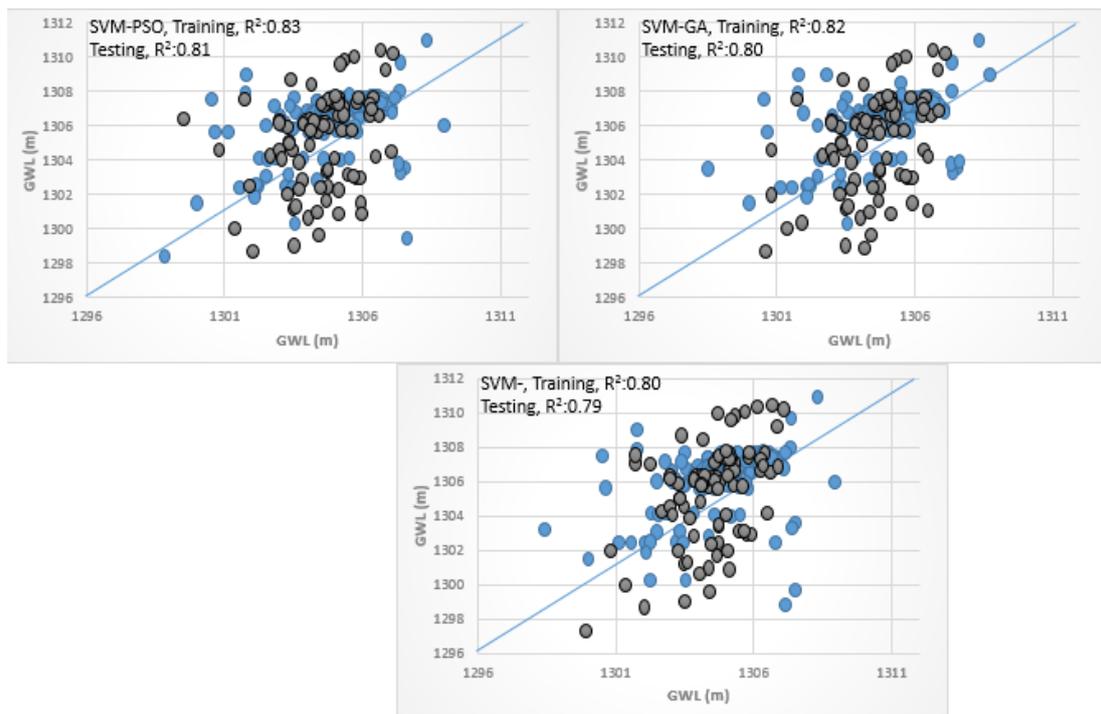

(c)

**Figure 7.** The scatter plots of exanimated soft computing models for predicting groundwater level (GWL), (**a**) piezometer 6, (**b**) piezometer 9, and (**c**) piezometer 10.

*3.5. Uncertainty Analysis of Soft Computing Models*

As stated in the aims of the current study, the uncertainty analysis of hybrid intelligence models is another major contribution and novelty of the present study. The same as the previous subsections, in this section the results of uncertainty analysis of hybrid models in selected three piezometers are provided and comparative evaluation between different hybrid models are presented. The hybrids of ANFIS, SVM, and ANN models with GOA, WA, KA, PSO, and GA are joined with the non-parametric Monte-Carlo Simulations (MCSs) to quantify the uncertainty of developed models in GWL simulations. The probability of model predictions in MCSs is considered as a degree of uncertainty of model results and demonstrates the probabilities in the GWL forecasting bands that enclosed the observed GWL inside these bounds of probability.

- Piezometer 6

In the trained hybrid models, the uncertainty in the model trained parameters and weights is the major source of uncertainty in model results. Here the effects of uncertainty in trained, optimization, and determination of parameters, and weights of intelligence developed hybrid models for piezometer 6 are presented. For training and testing stage, the uncertainty of the models results in piezometer 6 are provided in Figure 8a and in Table 6. The uncertainty results are quantified by the two indices of p and d and visualized by the uncertainty bounds of 95%. At first, the values of p show how many of the observed GWL values in the training and testing stages are positioned inside the 95% confidence bounds. Secondly, the d-factor as the measure of deviations should be small also. Figure 8 indicated that the highest and lowest d was obtained for SVM and ANFIS-GOA, respectively. Based on p and d indices, CSO had better performance than PSO, GA, KA, and WA. Results indicated that the standalone ANN, ANFIS, and SVM models had higher d and lower p than hybrid ANFIS, ANN, and SVM models that indicate higher uncertainty in the standalone model results. The overall comparison of the results indicated that the ANN model outperformed the SVM model. The d values of uncertainties of models in Table 6 show that in all of developed models for GWL the d value is lower than 1, that proves the superior tight bounds of developed models. The best results are derived by the ANFIS-GOA with d = 012 and p = 0.94 indicates that developed model 95% of observations are covered by the uncertainty bounds. The desired values for p in model uncertainty analysis have values greater than 80% [49].

- Piezometer 9

As presented in Table 6 and in Figure 8b, SVM-GOA and SVM had the lowest and highest d among SVM models. According to Table 6, GOA outperformed CSO and KA, but both algorithms were better than GA, PSO, and WA. The *p*-value of the standalone ANFIS model was increased by the optimization algorithms. GA provided lower performance in the optimization of ANN with p equal to 0.83 and d equal to 0.24, compared to WA, GOA, PSO, KA, CSO, and WA.

Again, the comparisons confirm the superiority of ANFIS-GOA in uncertainty verifications that have p = 0.94 and d = 0.16. As confirmed by these values of p in all of the developed models, all of them are satisfactory and the major part of GWL simulations are enclosed by the 95% prediction interval based on model prediction in Monte Carlo simulations. However, the d values that measure the average distance from upper and lower limits of prediction interval, for the ANFIS-GOA models are significantly and considerably smaller than those of all of the other models. In general, the benefits of ANFIS-GOA models over the other models is two-fold. At first, the GOA based models provide a more accurate prediction of GWL with fewer errors. Secondly, the confidence interval of ANFIS-GOA model results is much narrower and yet encloses almost the greatest percent of observation in MCSs.

- Piezometer 10

From Table 6, it was observed that ANFIS-GOA yielded the most dominant performance among other models. The weakest model in the optimization of the ANFIS model was ANFIS-GA with a p of 0.82 and d of 0.20. The ANN model provided better performance than the SVM model. The corresponding performance values of the SVM-GA model had p of 0.79 and d of 0.30. The standalone SVM model had the worst performance among other models.

However, general results indicated that the ANFIS-GOA has the best performance among other models. Figure 9 shows the coefficient of variation for different optimization algorithms. ANFIS-GOA had a lower coefficient of variation than other models and optimization algorithms. The worst results were for GA. In general, there are three main sources that generate the uncertainty of model outputs: the first one is the data and knowledge uncertainty, the second one is the parametric uncertainty due to unknown model parameters, and the third one is the structural uncertainty due to physical complexity of phenomena. The main contribution of the current paper is the uncertainty analysis of hybrid models prediction of GWL in the form of parametric uncertainty due to regulatory parameters and weights produced in the training stage of models.

**Table 6.** The results of uncertainty of soft computing models.

| Model | Piezometer 6 | | Piezometer 9 | | Piezometer 10 | |
|---|---|---|---|---|---|---|
| | p | d | p | d | p | d |
| ANFIS-GOA | 0.94 | 0.14 | 0.94 | 0.16 | 0.95 | 0.17 |
| ANN-GOA | 0.93 | 0.16 | 0.91 | 0.17 | 0.93 | 0.19 |
| SVM-GOA | 0.86 | 0.23 | 0.86 | 0.20 | 0.89 | 0.27 |
| ANFIS-CSO | 0.93 | 0.15 | 0.93 | 0.15 | 0.92 | 0.17 |
| ANN-CSO | 0.91 | 0.19 | 0.92 | 0.21 | 0.91 | 0.20 |
| SVM-CSO | 0.84 | 0.21 | 0.88 | 0.23 | 0.88 | 0.29 |
| ANFIS-KA | 0.90 | 0.15 | 0.92 | 0.17 | 0.89 | 0.18 |
| ANN-KA | 0.89 | 0.20 | 0.87 | 0.19 | 0.87 | 0.21 |
| SVM-KA | 0.86 | 0.21 | 0.89 | 0.19 | 0.86 | 0.29 |
| ANFIS-WA | 0.90 | 0.19 | 0.89 | 0.19 | 0.85 | 0.18 |
| ANN-WA | 0.86 | 0.23 | 0.84 | 0.24 | 0.84 | 0.19 |
| SVM-WA | 0.89 | 0.27 | 0.85 | 0.25 | 0.83 | 0.29 |
| ANFIS-PSO | 0.89 | 0.21 | 0.86 | 0.19 | 0.84 | 0.19 |
| ANN-PSO | 0.84 | 0.25 | 0.85 | 0.20 | 0.82 | 0.21 |
| SVM-PSO | 0.84 | 0.27 | 0.84 | 0.24 | 0.81 | 0.31 |
| ANFIS-GA | 0.87 | 0.20 | 0.83 | 0.24 | 0.82 | 0.20 |
| ANN-GA | 0.84 | 0.27 | 0.86 | 0.25 | 0.80 | 0.25 |
| SVM-GA | 0.80 | 0.32 | 0.89 | 0.29 | 0.79 | 0.30 |
| ANFIS | 0.85 | 0.20 | 0.87 | 0.24 | 0.78 | 0.20 |
| ANN | 0.82 | 0.28 | 0.83 | 0.24 | 0.77 | 0.27 |
| SVM | 0.80 | 0.35 | 0.82 | 0.29 | 0.76 | 0.33 |

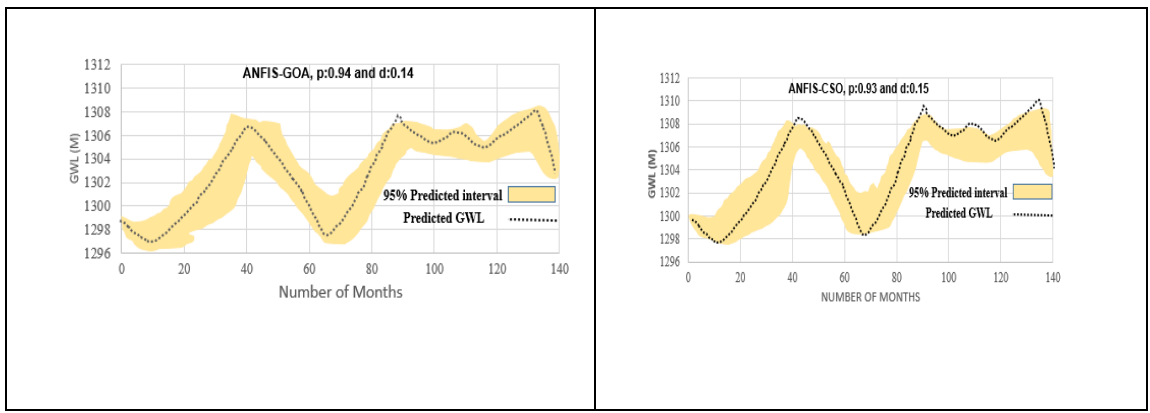

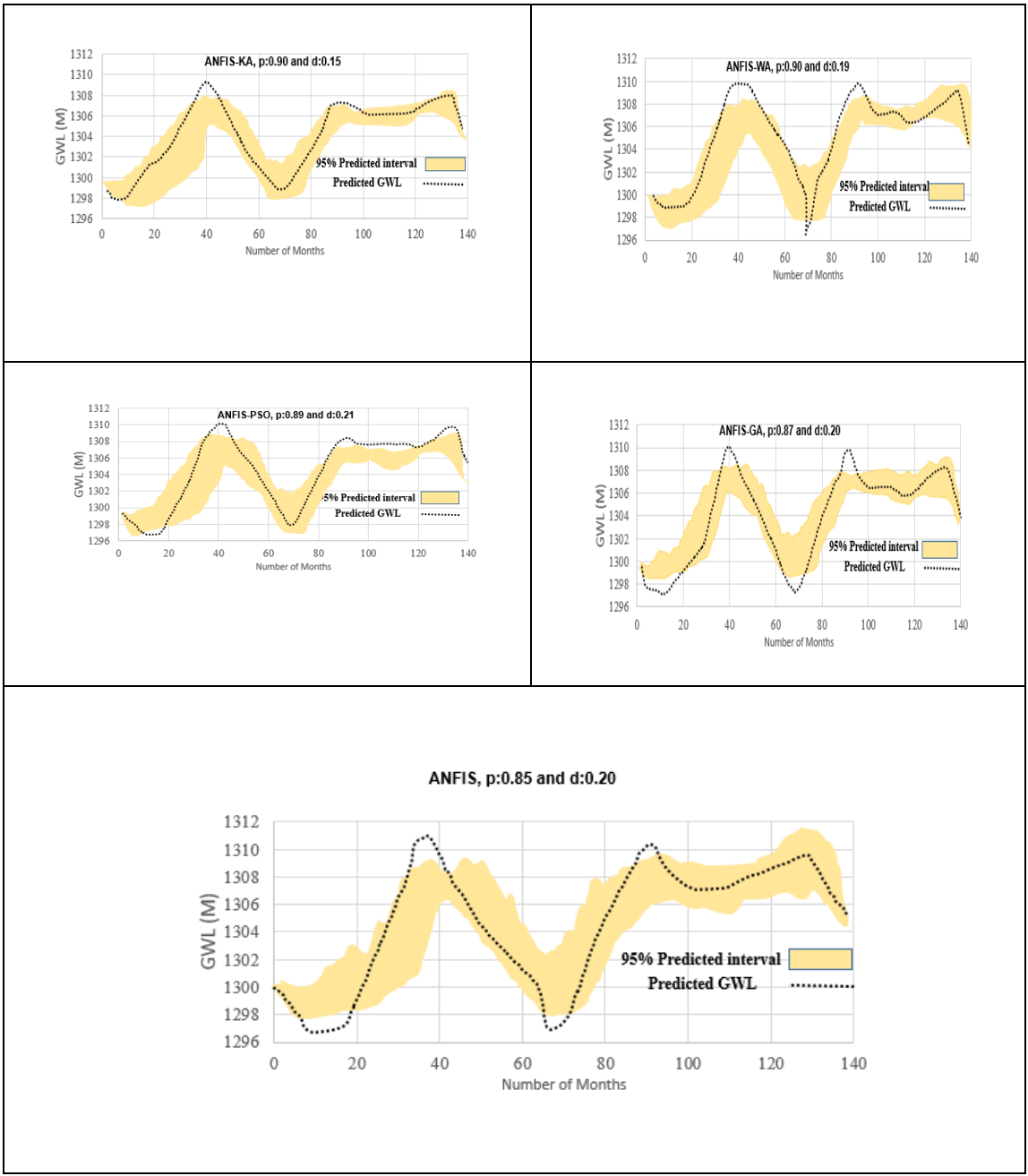

(a)

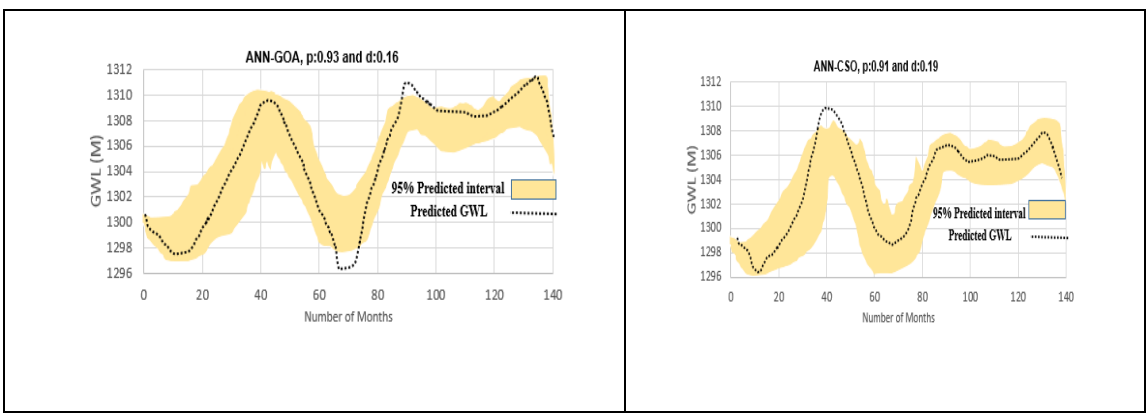

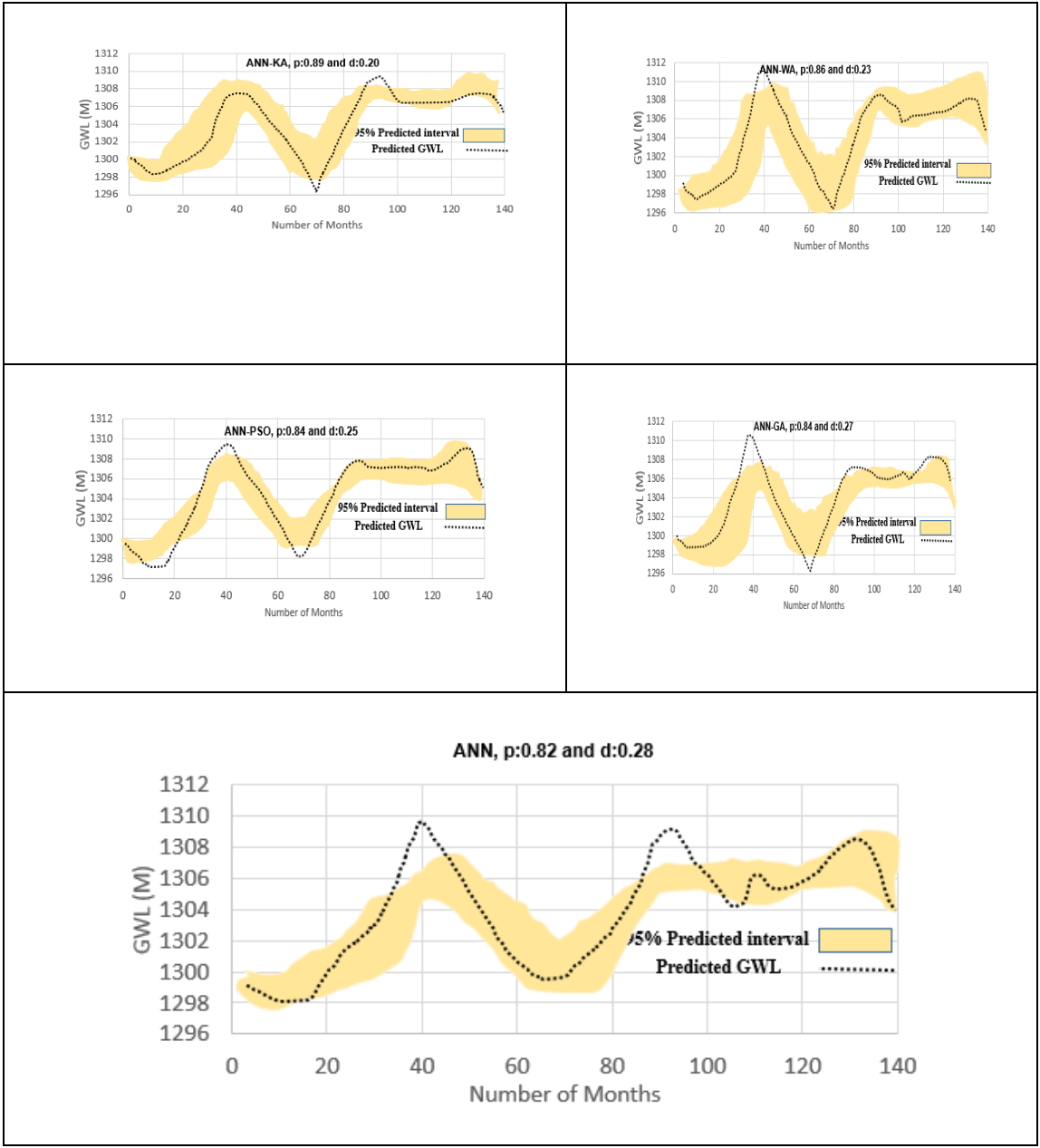

(b)

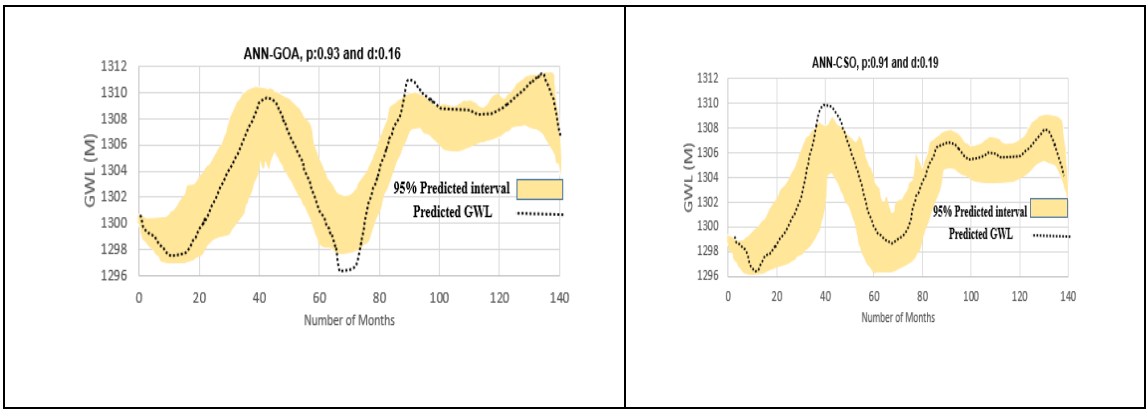

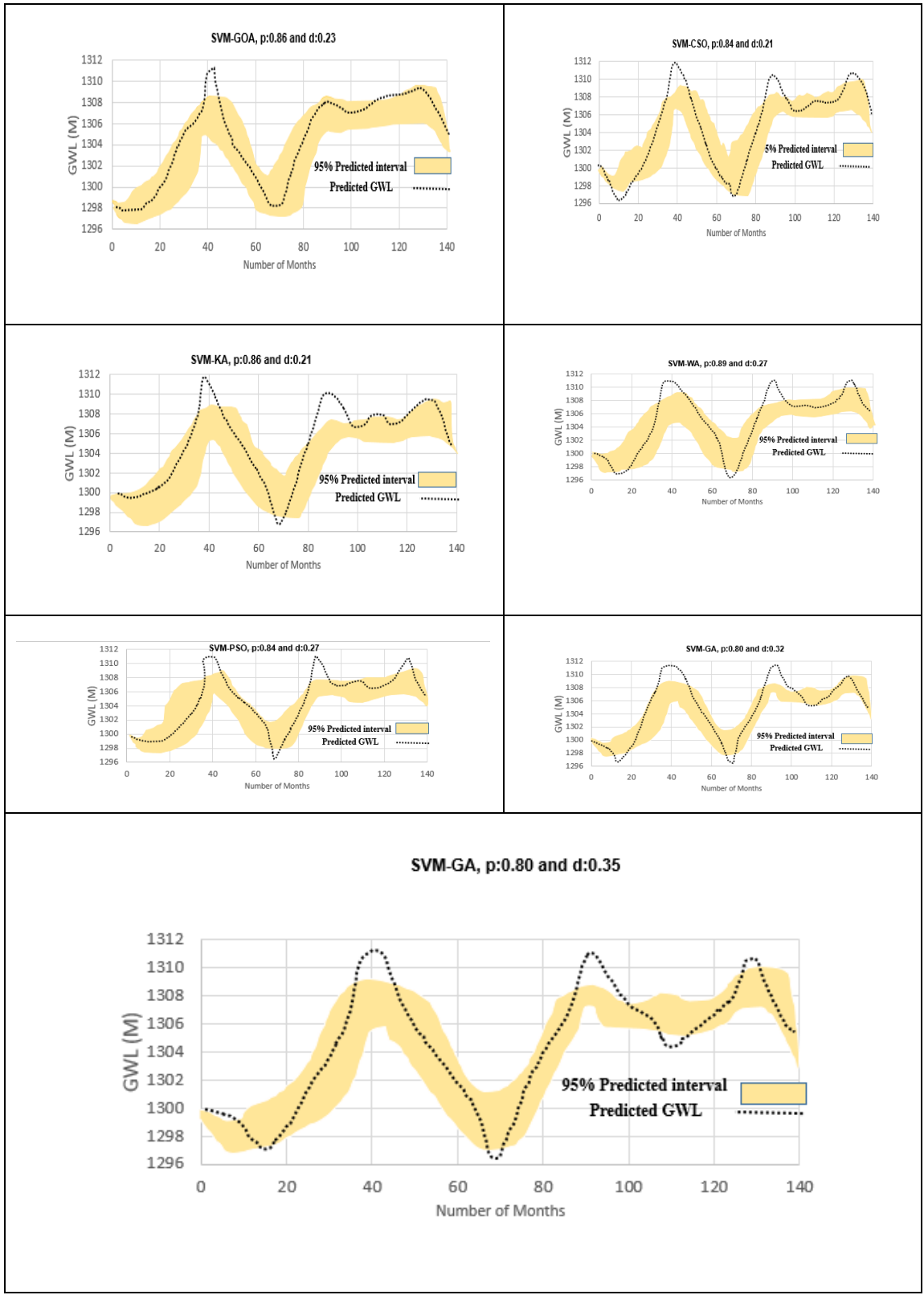

**Figure 8.** Computed uncertainty bound for piezometer 6; (**a**) ANFIS, (**b**) ANN, (**c**) SVM.

*3.6. Spatiotemporal Variation of GWL*

The previous section indicated that the GOA improved the performance of ANN, ANFIS, and SVM models. The results indicated that the GOA had better performance than other optimization algorithms. As shown in Figure 9, the hybrid GOA models (ANFIS-GOA, ANN-GOA, and SVM-GOA) have low variation coefficients in modeling.

Most literature reviews revealed only a few quantity comparisons. Furthermore, they did not include the spatiotemporal variation of GWL. In this section, the latitude, longitude, H(t-1), H (t-2), H (t-3), H (t-4), and H (t-5), hydraulic conductivity (HC), and specific yield of nine observed wells (well 6, 9, 10, 24, 11, 4, 7, 8, and 1) were used to provide the spatiotemporal variation of GWL for different months. The Ardebil plain is a heterogeneous aquifer. Thus, the hydraulic conductivity and specific yield spatially vary in the Ardebil plain. HC is a measure of a material's capacity to transmit water. The specific yield is defined as the ratio of the volume of water that an aquifer will yield by gravity to the total volume of the aquifer. A pumping test method was used to obtain the value of the hydraulic conductivity and specific yield. Figure 10 shows the measured hydraulic conductivity and specific yield for the Ardebil plain. In this section, the ANFIS, ANN, and SVM models with the best algorithm (GOA) were used to provide the spatiotemporal variation of GWL. The difference between estimated GWL models and observed GWL was computed for all months of years. The RMSE was used as an error function to compare the estimated data with the observed data. From Figure 11, it was clear that the ANFIS-GOA provided more accurate estimation than ANN-GOA and SVM-GOA. It was clear that the RMSE of ANFIS-GOA varied from white (1.2 m) to dark blue (2.2), while the RMSE of ANN-GOA and SVM-GOA varied from 1.7 (yellow) to 2.7 m (light green). Thus, results indicated that ANFIS-GOA has higher accuracy for the heterogeneous aquifers. The heterogeneous aquifers are considered as complex hydraulic systems because their hydraulic parameters vary spatially and temporally. Additionally, the climate parameters, such as temperature and rainfall, can increase the complexity of prediction of GWL for heterogeneous aquifers.

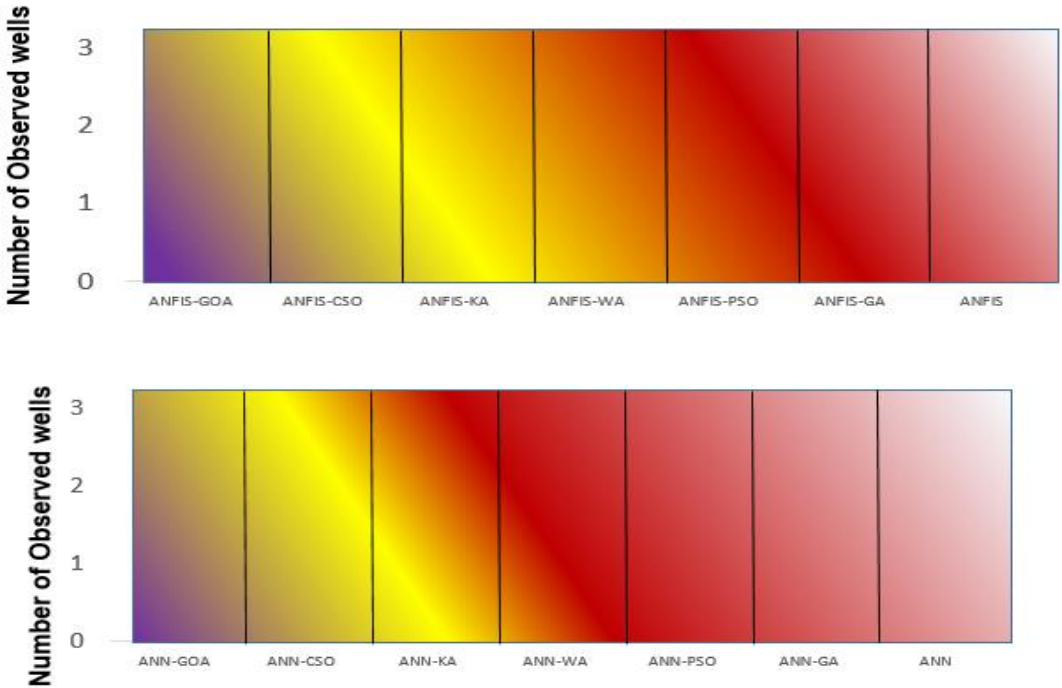

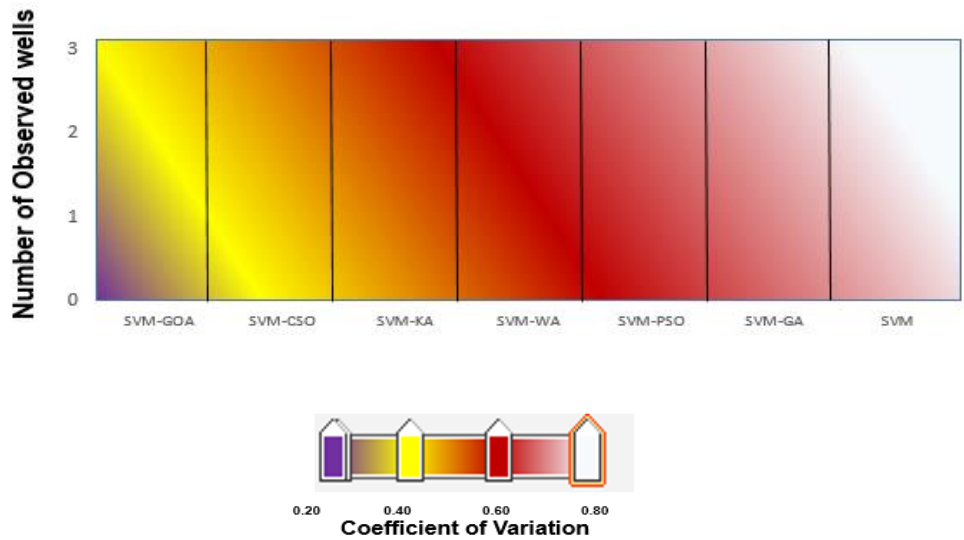

**Figure 9.** The map of variations coefficient of different models for 100 random runs of objective function.

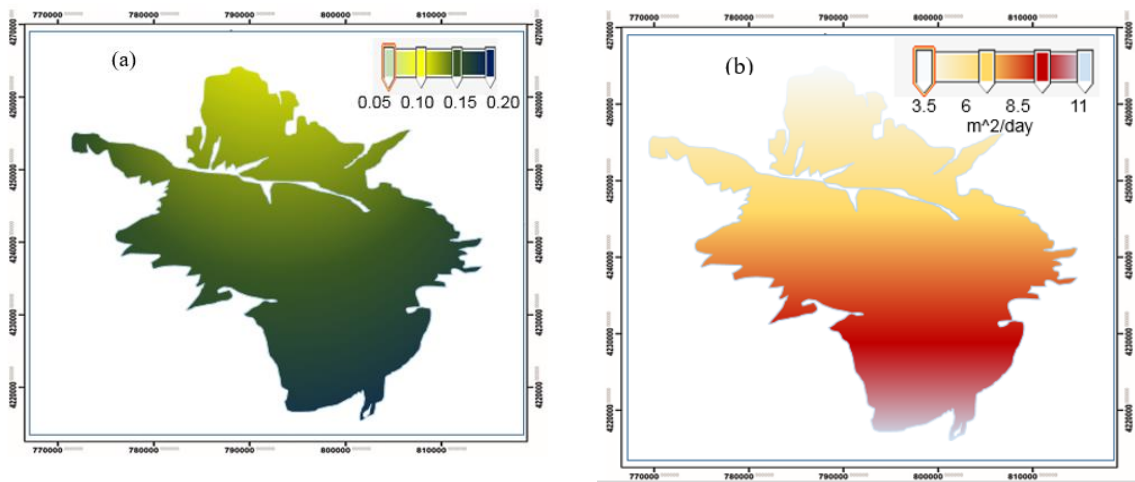

**Figure 10.** (**a**) Spatial specific yield and (**b**) hydraulic conductivity.

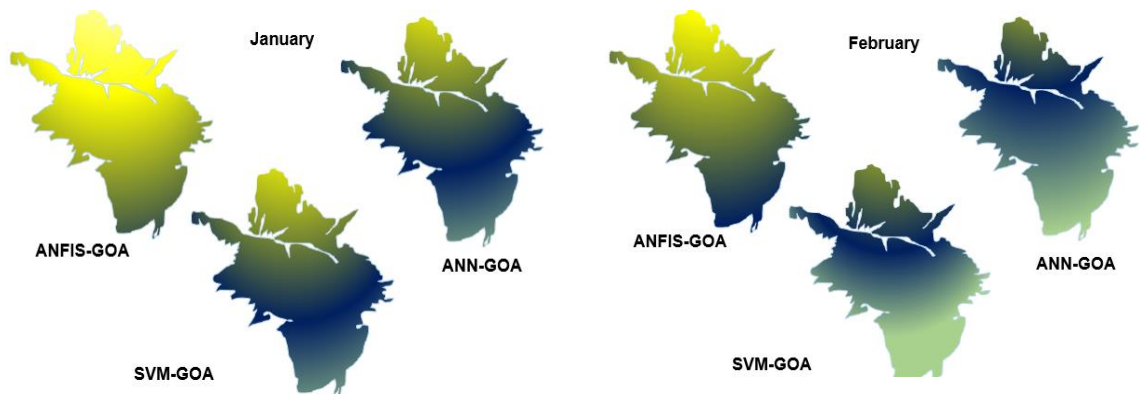

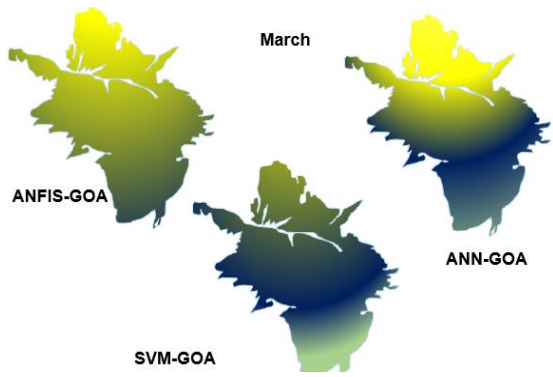
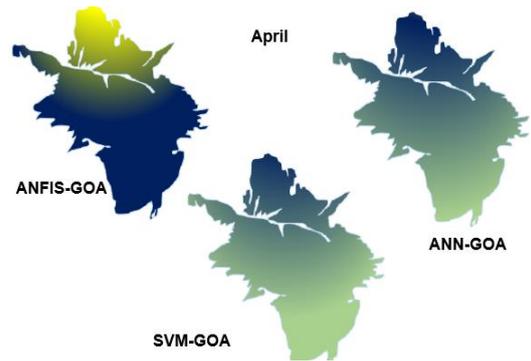
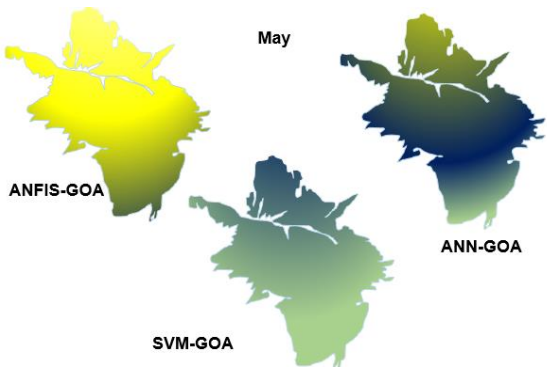
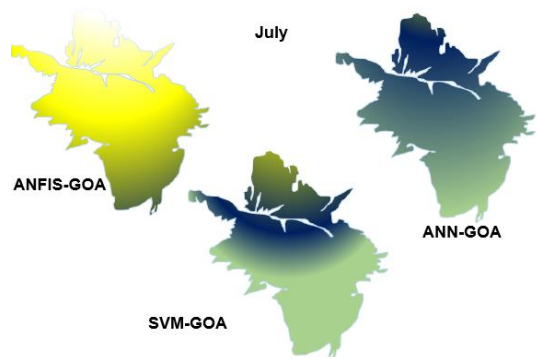
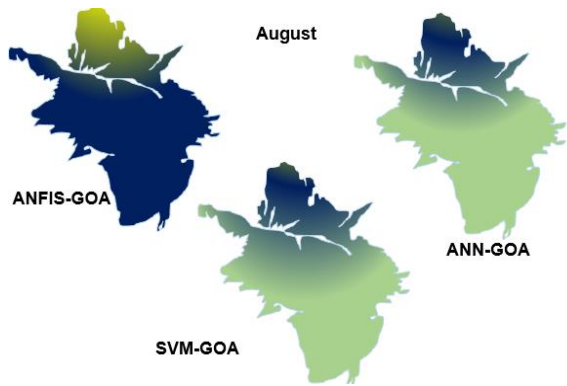
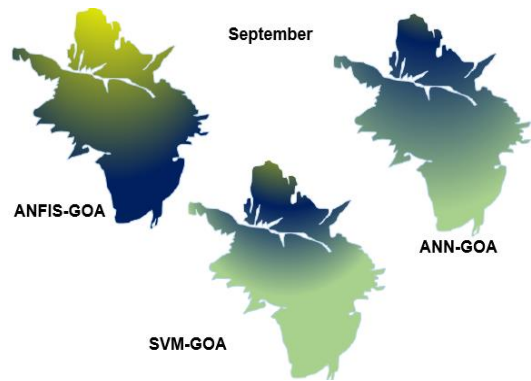
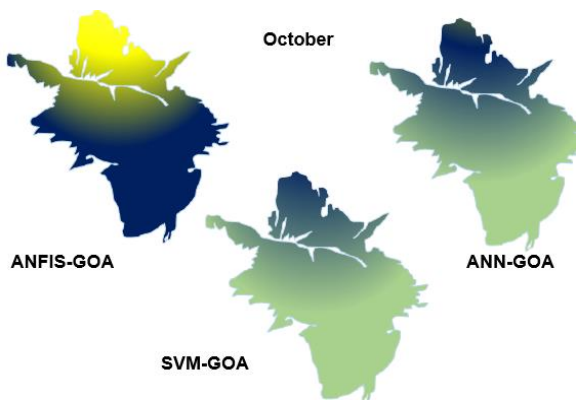
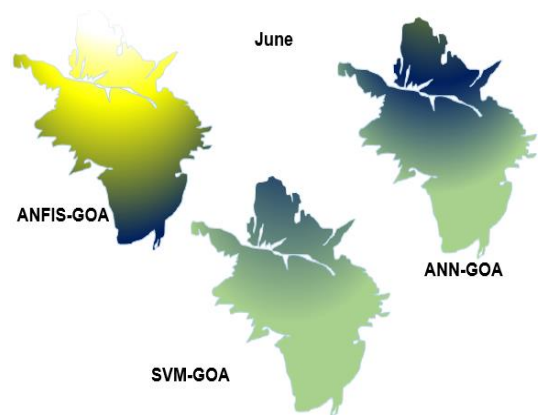

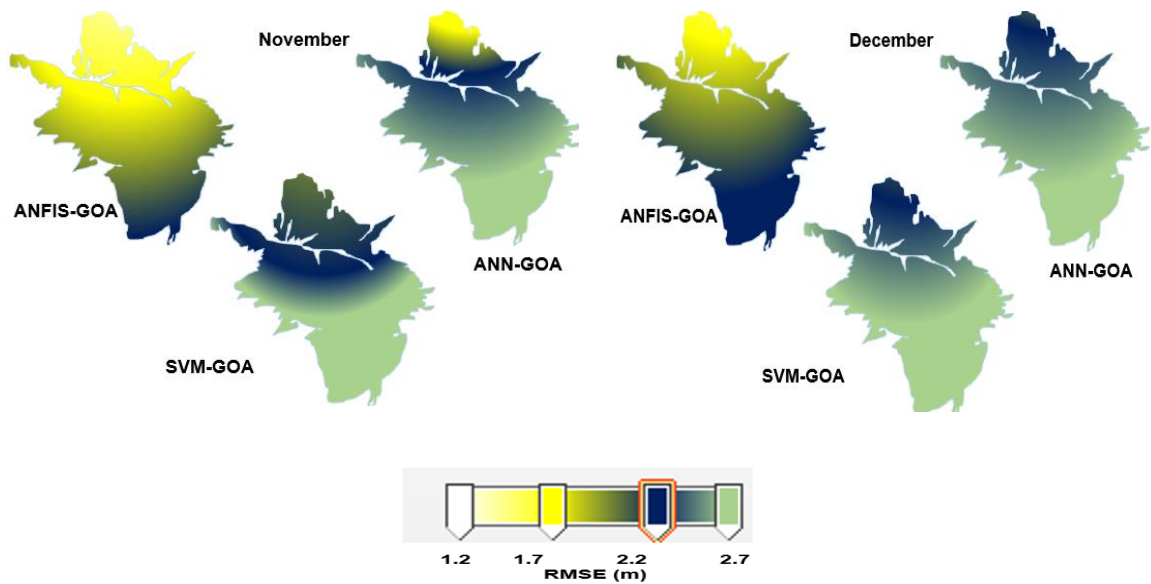

**Figure 11.** The spatial and temporal variation of GWL.

## 4. Conclusions

In this study, the ANFIS, ANN, and SVM models were used to predict groundwater level. The GOA, CSO, GA, PSO, WA, and KA were used to fine-tune and integrate with the ANN, SVM, and ANFIS models. Three piezometers (6, 9, and 10) in the Ardebil plain were considered as a case study for the GWL investigation. The input combinations of time series (up to 12-month lag) were reduced using principal component analysis (PCA). For the testing phase and piezometer 6 ANFIS-GOA indicated a value of RMSE: 1.21, MAE: 0.878, NSE: 0.93, and PBIAS: 0.15 which reflected better performance than the other models. The $R^2$ values were found to vary in the range of 0.84–0.94 and 0.79–0.91 for the ANN (hybrid ANN models and based ANN model) and SVM models (hybrid SVM models and based SVM model), respectively. The results indicated that the SVM model had the lowest $R^2$ among other models. It was observed that the ANFIS-GOA yielded the most dominant performance among other models. From uncertainty analysis, the weakest model in the optimization of the ANFIS model was ANFIS-GA with a $p = 0.87$ and $d = 0.21$. However, general results indicated that the ANFIS-GOA had better performance than other models. Additionally, the results of spatiotemporal variations maps of GWL showed that ANFIS-GOA has high accuracy for the heterogeneous Ardebil aquifer. Future studies can evaluate the accuracy of these models under climate change conditions. The climate parameters such as temperature and rainfall can be simulated for future periods. Then, these parameters can be used as input to the models to simulate GWL for the future periods.

**Author Contributions:** Conceptualization, A.S., and M.E.; methodology, A.S., and M.E.; writing—review and editing, A.S., M.E., V.P.S., and A.M..; validation; A.M., A.S., and M.E.; supervision, V.P.S.; Funding acquisition, A.M. All authors have read and agreed to the published version of the manuscript.

**Funding:** This work is supported by the Hungarian State and the European Union under the EFOP-3.6.1-16-2016-00010 project and the 2017-1.3.1-VKE-2017-00025 project.

**Conflicts of Interest**: The authors declare no conflict of interest.

**Acknowledgments:** Support of the Hungarian State and the European Union under the EFOP-3.6.1-16-2016-00010 project and the 2017-1.3.1-VKE-2017-00025 project is acknowledged. We also acknowledge the support of the German Research Foundation (DFG) and the Bauhaus-Universität Weimar within the Open-Access Publishing Programme.

# References

1. Sattari, M. T.; Mirabbasi, R.; Sushab, R. S.; Abraham, J. Prediction of Groundwater Level in Ardebil Plain Using Support Vector Regression and M5 Tree Model. *Groundwater* **2018**. https://doi.org/10.1111/gwat.12620.Jeong,
2. J.; Park, E. Comparative applications of data-driven models representing water table fluctuations. *J. Hydrol*. **2019**, *572*, 261–273, doi:10.1016/j.jhydrol.2019.02.051.
3. Alizamir, M.; Kisi, O.; Zounemat-Kermani, M. Modelling long-term groundwater fluctuations by extreme learning machine using hydro-climatic data. *Hydrol. Sci. J.* **2018**, *63*, 63–73, doi:10.1080/02626667.2017.1410891.
4. Yoon, H.; Kim, Y.; Lee, S.H.; Ha, K. Influence of the range of data on the performance of ANN-and SVM-based time series models for reproducing groundwater level observations. *Acque Sotter. Ital. J. Groundwater*. **2019**, doi:10.7343/as-2019-376.
5. Mohanty, S.; Jha, M.K.; Kumar, A.; Sudheer, K.P. Artificial neural network modeling for groundwater level forecasting in a river island of eastern India. *Water Resour. Manag.* **2010**, *24*, 1845–1865, doi:10.1007/s11269-009-9527-x.
6. Natarajan, N.; Sudheer, C. Groundwater level forecasting using soft computing techniques. *Neural Comput. Appl.* **2019**, 1–18, doi:10.1007/s00521-019-04234-5.
7. Lee, S.; Lee, K.K.; Yoon, H. Using artificial neural network models for groundwater level forecasting and assessment of the relative impacts of influencing factors. *Hydrogeol. J.* **2019**, *27*, 567–579, doi:10.1007/s10040-018-1866-3.
8. Khan, U.T.; Valeo, C. Dissolved oxygen prediction using a possibility theory based fuzzy neural network. *Hydrol. Earth Syst. Sci*. **2016**, *20*, 2267–2293, doi:10.5194/hess-20-2267-2016.
9. Jeihouni, E.; Eslamian, S.; Mohammadi, M.; Zareian, M.J. Simulation of groundwater level fluctuations in response to main climate parameters using a wavelet–ANN hybrid technique for the Shabestar Plain, Iran. *Environ. Earth Sci*. **2019**, *78*, 293, doi:10.1007/s12665-019-8283-3.
10. Alian, S.; Mayer, A.; Maclean, A.; Watkins, D.; Mirchi, A. Spatiotemporal Dimensions of Water Stress Accounting: Incorporating Groundwater–Surface Water Interactions and Ecological Thresholds. Environ. Sci. Technol. **2019**, *53*, 2316–2323, doi:10.1021/acs.est.8b04804.
11. Jalalkamali, A.; Sedghi, H.; Manshouri, M. Monthly groundwater level prediction using ANN and neuro-fuzzy models: A case study on Kerman plain, Iran. *J. Hydroinformatics*. **2010**, *13*, 867–876, doi:10.2166/hydro.2010.034.
12. Trichakis, I.C.; Nikolos, I.K.; Karatzas, G.P. Artificial neural network (ANN) based modeling for karstic groundwater level simulation. *Water Resour. Manag.* **2011**, *25*, 1143–1152, doi:10.1007/s11269-010-9628-6.
13. Fallah-Mehdipour, E.; Haddad, O.B.; Mariño, M.A. Prediction and simulation of monthly groundwater levels by genetic programming. J Hydro-Environment. Res. **2013**, *7*, 253–260, doi:10.1016/j.jher.2013.03.005.
14. Moosavi, V.; Vafakhah, M.; Shirmohammadi, B.; Behnia, N. A wavelet-ANFIS hybrid model for groundwater level forecasting for different prediction periods. *Water Resour. Manag.* **2013**, *27*, 1301–1321, doi:10.1007/s11269-012-0239-2.
15. Emamgholizadeh, S.; Moslemi, K.; Karami, G. Prediction the Groundwater Level of Bastam Plain (Iran) by Artificial Neural Network (ANN) and Adaptive Neuro-Fuzzy Inference System (ANFIS). *Water Resour. Manag*. **2014**, *28*, 5433–5446, doi:10.1007/s11269-014-0810-0.
16. Suryanarayana, C.; Sudheer, C.; Mahammood, V.; Panigrahi, B.K. An integrated wavelet-support vector machine for groundwater level prediction in Visakhapatnam, India. *Neurocomputing*. **2014**, *145*, 324–335, doi:10.1016/j.neucom.2014.05.026.
17. Mohanty, S.; Jha, M.K.; Raul, S.K.; Panda, R.K.; Sudheer, K.P. Using artificial neural network approach for simultaneous forecasting of weekly groundwater levels at multiple sites. *Water Resour. Manag*. **2015**, *29*, 5521–5532, doi:10.1007/s11269-015-1132-6.
18. Yoon, H.; Hyun, Y.; Ha, K.; Lee, K.K.; Kim, G.B. A method to improve the stability and accuracy of ANN- and SVM-based time series models for long-term groundwater level predictions. *Comput. Geosci*. **2016**, *90*, 144–155, doi:10.1016/j.cageo.2016.03.002.
19. Zhou, T.; Wang, F.; Yang, Z. Comparative analysis of ANN and SVM models combined with wavelet preprocess for groundwater depth prediction. *Water* **2017**, *9*, 781, doi:10.3390/w9100781.
20. Choubin, B.; Malekian, A. Combined gamma and M-test-based ANN and ARIMA models for groundwater fluctuation forecasting in semiarid regions. *Environ. Earth Sci.* **2017**, *76*, 538, doi:10.1007/s12665-017-6870-8.

21. Das, U.K.; Roy, P.; Ghose, D.K. Modeling water table depth using adaptive Neuro-Fuzzy Inference System. *ISH J. Hydraul. Eng.* **2019**, *25*, 291–297, doi:10.1080/09715010.2017.1420497.
22. Hadipour, A.; Khoshand, A.; Rahimi, K.; Kamalan, H.R. Groundwater Level Forecasting by Application of Artificial Neural Network Approach: A Case Study in Qom Plain, Iran. *J. Hydrosci. Environ.* **2019**, *3*, 30–34, doi:10.22111/jhe.2019.4972.
23. Jalalkamali, A.; Jalalkamali, N. Groundwater modeling using hybrid of artificial neural network with genetic algorithm. *Afr. J. Agric. Res*. **2011**, *6*, 5775–5784, doi:10.5897/AJAR11.1892.
24. Mathur, S. Groundwater level forecasting using SVM-PSO. *Int. J. Hydrol. Sci. Technol*. **2012**, *2*, 202–218, doi:10.1504/IJHST.2012.047432.
25. Hosseini, Z.; Gharechelou, S.; Nakhaei, M.; Gharechelou, S. Optimal design of BP algorithm by ACO R model for groundwater-level forecasting: A case study on Shabestar plain, Iran. *Arab. J. Geosci*. **2016**, *9*, 436, doi:10.1007/s12517-016-2454-2.
26. Zare, M.; Koch, M. Groundwater level fluctuations simulation and prediction by ANFIS-and hybrid Wavelet-ANFIS/Fuzzy C-Means (FCM) clustering models: Application to the Miandarband plain. *J. Hydro-Environ. Res*. **2018**, *18*, 63–76, doi:10.1016/j.jher.2017.11.004.
27. Balavalikar, S.; Nayak, P.; Shenoy, N.; Nayak, K. Particle swarm optimization based artificial neural network model for forecasting groundwater level in Udupi district. In *Proceedings of the AIP Conference*; AIP Elsevier. New York. USA: 2018. doi:10.1063/1.5031983.
28. Malekzadeh, M.; Kardar, S.; Saeb, K.; Shabanlou, S.; Taghavi, L. A Novel Approach for Prediction of Monthly Ground Water Level Using a Hybrid Wavelet and Non-Tuned Self-Adaptive Machine Learning Model. *Water Resour. Manag.* **2019**, *33*, 1609–1628, doi:10.1007/s11269-019-2193-8.
29. Mirjalili, S.Z.; Mirjalili, S.; Saremi, S.; Faris, H.; Aljarah, I. Grasshopper optimization algorithm for multi-objective optimization problems. *Appl. Intell*. **2018**, *48*, 805–820, doi:10.1007/s10489-017-1019-8.
30. Arora, S.; Anand, P. Chaotic Grasshopper Optimization Algorithm for Global Optimization. *Neural Comput. Appl.* **2019**. https://doi.org/10.1007/s00521-018-3343-2.
31. Alizadeh, Z.; Yazdi, J.; Kim, J.; Al-Shamiri, A. Assessment of Machine Learning Techniques for Monthly Flow Prediction. *Water* **2018**, *10*, 1676, doi:10.3390/w10111676.
32. Moayedi, H.; Gör, M.; Lyu, Z.; Bui, D.T. Herding Behaviors of Grasshopper and Harris hawk for Hybridizing the Neural Network in Predicting the Soil Compression Coefficient. *Meas. J. Int. Meas. Confed*. **2019a**, 107389, doi:10.1016/j.measurement.2019.107389.
33. Gampa, S.R.; Jasthi, K.; Goli, P.; Das, D.; Bansal, R.C. Grasshopper optimization algorithm based two stage fuzzy multiobjective approach for optimum sizing and placement of distributed generations, shunt capacitors and electric vehicle charging stations. *J. Energy Storage.* **2020**, *27*, 101117, doi:10.1016/j.est.2019.101117.
34. Moayedi, H.; Kalantar, B.; Foong, L.K.; Tien Bui, D.; Motevalli, A. Application of three metaheuristic techniques in simulation of concrete slump. *Appl. Sci*. **2019b**, *9*, 4340, doi:10.3390/app9204340.
35. Kumar, A.; Kumar, P.; Singh, V.K. Evaluating Different Machine Learning Models for Runoff and Suspended Sediment Simulation. *Water Resour. Manag*. **2019**, *33*, 1217–1231, doi:10.1007/s11269-018-2178-z.
36. Khosravi, K.; Daggupati, P.; Alami, M.T.; Awadh, S.M.; Ghareb, M.I.; Panahi, M.; ThaiPham, B.; Rezaei, F.; Qi, C.; Yaseen, Z.M. Meteorological data mining and hybrid data-intelligence models for reference evaporation simulation: A case study in Iraq. *Comput. Electron. Agric*. **2019**, *167*, 105041, doi:10.1016/j.compag.2019.105041.
37. Kisi, O.; Yaseen, Z.M. The potential of hybrid evolutionary fuzzy intelligence model for suspended sediment concentration prediction. *Catena* **2019**, *174*, 11–23, doi:10.1016/j.catena.2018.10.047.
38. Dubdub, I.; Rushd, S.; AlYaari, M.; Ahmed, E. Application of Artificial Neural Network to Model the Pressure Losses in the Water-Assisted Pipeline Transportation of Heavy Oil. Proceedings of the SPE Middle East Oil and Gas Show and Conference. *Soc. Pet. Eng*. **2019**, doi:10.2118/194742-MS.
39. Moghaddam, H.K.; Moghaddam, H.K.; Kivi, Z.R.; Bahreinimotlagh, M.; Alizadeh, M.J. Developing comparative mathematic models, BN and ANN for forecasting of groundwater levels. *Groundw. Sustain. Dev*. **2019**, 100237, doi:10.1016/j.gsd.2019.100237.
40. Fan, J.; Wang, X.; Wu, L.; Zhou, H.; Zhang, F.; Yu, X.; Lu, X.; Xiang, Y. Comparison of Support Vector Machine and Extreme Gradient Boosting for predicting daily global solar radiation using temperature and precipitation in humid subtropical climates: A case study in China. *Energy Convers. Manag*. **2018**, *164*, 102–111, doi:10.1016/j.enconman.2018.02.087.

41. Pour, S.H.; Shahid, S.; Chung, E.S.; Wang, X.J. Model output statistics downscaling using support vector machine for the projection of spatial and temporal changes in rainfall of Bangladesh. *Atmos. Res*. **2018**, *213*, 149–162, doi:10.1016/j.atmosres.2018.06.006.
42. Pham, B.T.; Bui, D.T.; Prakash, I. Bagging based Support Vector Machines for spatial prediction of landslides. *Environ. Earth Sci*. **2018**, *77*, 146, doi:10.1007/s12665-018-7268-y.
43. Deo, R.C.; Salcedo-Sanz, S.; Carro-Calvo, L.; Saavedra-Moreno, B. Drought prediction with standardized precipitation and evapotranspiration index and support vector regression models. In *Integrating Disaster Science and Management*, 1st ed.; Elsevier: Amsterdam, The Netherlands, 2018; pp. 151–174. doi:10.1016/B978-0-12-812056-9.00010-5.
44. Mafarja, M.; Aljarah, I.; Faris, H.; Hammouri, A.I.; Ala'M, A.Z.; Mirjalili, S. Binary grasshopper optimisation algorithm approaches for feature selection problems. *Expert Syst. Appl*. **2019**, *117*, 267–286, doi:10.1016/j.eswa.2018.09.015.
45. Mehrabian, A.R.; Lucas, C. A novel numerical optimization algorithm inspired from weed colonization. *Ecol. Inform*. **2006**, *1*, 355–366, doi:10.1016/j.ecoinf.2006.07.003.
46. Chandirasekaran, D.; Jayabarathi, T. Cat swarm algorithm in wireless sensor networks for optimized cluster head selection: A real time approach. *Cluster Comput*. **2019**, *22*, 11351–11361, doi:10.1007/s10586-017-1392-4.
47. Karpenko, A.P.; Leshchev, I.A. Advanced Cat Swarm Optimization Algorithm in Group Robotics Problem. *Procedia Comput. Sci*. **2019**, *150*, 95–101, doi:10.1016/j.procs.2019.02.020.
48. Ramezani, F. Solving Data Clustering Problems using Chaos Embedded Cat Swarm Optimization. *J. Adv. Comput. Res.* **2019**, *10*, 1–10.
49. Orouskhani, M.; Shi, D. Fuzzy adaptive cat swarm algorithm and Borda method for solving dynamic multi-objective problems. *Expert Syst*. **2018**, *35*, e12286, doi:10.1111/exsy.12286.
50. Pradhan, P.M.; Panda, G. Solving multiobjective problems using cat swarm optimization. *Expert Syst. Appl*. **2012**, *39*, 2956–2964, doi:10.1016/j.eswa.2011.08.157.
51. Chu, S.C.; Tsai, P.W.; Pan, J.S. Cat swarm optimization. In *Proceedings of the Pacific Rim International Conference on Artificial Intelligence*; Springer: Berlin/Heidelberg, Germany, 2006; pp. 854–858. doi:10.1007/11801603_94.
52. Saha, S.K.; Ghoshal, S.P.; Kar, R.; Mandal, D. Cat swarm optimization algorithm for optimal linear phase FIR filter design. *ISA Trans*. **2013**, *52*, 781–794, doi:10.1016/j.isatra.2013.07.009.
53. Kennedy, J.; Eberhart, R.C. A discrete binary version of the particle swarm algorithm. Proceedings of 1997 IEEE International conference on systems, man, and cybernetics. *Comput. Cybern. Simul. IEEE* **1997**, 4104–4108, doi:10.1109/icsmc.1997.637339.
54. Wang, G. G.; Gandomi, A. H.; Alavi, A. H.; Gong, D. A Comprehensive Review of Krill Herd Algorithm: Variants, Hybrids and Applications. *Artif. Intell. Rev.* 2019. https://doi.org/10.1007/s10462-017-9559-1.
55. Abualigah, L.M.; Khader, A.T.; Hanandeh, E.S. A combination of objective functions and hybrid Krill herd algorithm for text document clustering analysis. *Eng. Appl. Artif. Intell*. **2018**, *73*, 111–125, doi:10.1016/j.engappai.2018.05.003.
56. Asteris, P.G.; Nozhati, S.; Nikoo, M.; Cavaleri, L.; Nikoo, M. Krill herd algorithm-based neural network in structural seismic reliability evaluation. *Mech. Adv. Mater. Struct*. **2019**, *26*, 1146–1153, doi:10.1080/15376494.2018.1430874.
57. Lu, C.; Feng, J.; Liu, W.; Lin, Z.; Yan, S. Tensor robust principal component analysis with a new tensor nuclear norm. *IEEE Trans. Pattern Anal. Mach. Intell.* **2019**, doi:10.1109/TPAMI.2019.2891760.
58. Priyadarshi, D.; Paul, K.K. Optimisation of biodiesel production using Taguchi model. *Waste Biomass Valori*. **2019**, *10*, 1547–1559, doi:10.1007/s12649-017-0158-9.
59. Yen, H.; Wang, X.; Fontane, D.G.; Harmel, R.D.; Arabi, M. A Framework for Propagation of Uncertainty Contributed by Parameterization, Input Data, Model Structure, and Calibration/Validation Data in Watershed Modeling. *Environ. Model. Softw*. **2014**, doi:10.1016/j.envsoft.2014.01.004.
60. Youcai, Z.; Sheng, H. Pollution Characteristics of Industrial Construction and Demolition Waste. In *Pollution Control and Resource Recovery*; 2017; USA, New York. doi:10.1016/b978-0-12-811754-5.00004-x.